\documentclass[preprint,12pt]{elsarticle}

\usepackage{amssymb}
\usepackage{pdfpages}
\usepackage[draft]{changes}

\begin{document}

\begin{frontmatter}



\title{Kinematics reconstruction in solenoidal spectrometers operated in active target mode}


\author[IGFAE]{Yassid Ayyad}
\author[MSU,FRIB]{Adam K. Anthony}
\author[MSU,FRIB]{Daniel Bazin}
\author[SUST]{Jie Chen}
\author[MSU,FRIB]{Wolfgang Mittig}
\author[ARG]{Ben P. Kay}
\author[MAN]{David K. Sharp}
\author[MSU,FRIB]{Juan Carlos Zamora}

\affiliation[IGFAE]{IGFAE, Universidade de Santiago de Compostela, E-15782, Santiago de Compostela, Spain}
\affiliation[MSU]{Department of Physics and Astronomy, Michigan State University. East Lansing, MI, 48824,USA}
\affiliation[FRIB]{Facility for Rare Isotope Beams, Michigan State University, East Lansing, MI 48824, USA}
\affiliation[SUST]{organization={College of Science, Southern University of Science and Technology},  city={Shenzhen}, postcode={518055}, province={Guangdong},country={China}}
\affiliation[ARG]{Physics Division, Argonne National Laboratory, Argonne, IL 60439, USA}
\affiliation[MAN]{organization={School of Physics and Astronomy, University of Manchester},city={Manchester},postcode={M13 9PL}, country={UK} }

\begin{abstract}
We discuss the reconstruction of low-energy nuclear reaction kinematics from charged-particle tracks in solenoidal spectrometers working in Active Target Time Projection Chamber mode. In this operation mode, reaction products are tracked within the active gas medium of the Active Target with a three dimensional space point cloud. We have inferred the reaction kinematics from the point cloud using an algorithm based on a linear quadratic estimator (Kalman filter). The performance of this algorithm has been evaluated using experimental data from nuclear reactions measured with the Active Target Time Projection Chamber (AT-TPC) detector. 
\end{abstract}




\end{frontmatter}


\section{Introduction}
\label{sec:sec1}

The recent advances in radioactive (or exotic) isotope production has laid the foundations to redefine the goals of the modern low-energy nuclear physics by providing access to astonishing properties of isospin-imbalanced nuclear matter~\cite{osti_1296778}. Arguably, some of the most prominent features strongly affected by the dramatic reorganization of nuclear matter at the limits of stability include: the evolution of the shell structure, collective phenomena involving oscillations, rotations and vibrations, the coexistence of nuclear shapes at close energies, molecule-like clustering, among many others. From an experimental stand-point, the path to study such phenomena can be accessed via controlled nuclear reactions in inverse kinematics where the heavy radioactive nucleus is accelerated and impinges on a light target~\cite{Shapira1985}. The selection of a suitable reaction mechanism as a probe depends very much on the phenomena to be investigated. One of the most powerful tools, for the cases explored here, are direct reactions, such as inelastic scattering or nucleon transfer reactions~\cite{Wimmer_2018,Catford2014}. These peripheral reactions can be used to obtain single-particle or collective properties of the nucleus using relatively simple observables. Conventional direct-reaction experiments usually require minimum intensities of the order of 10$^{4-5}$ particles per second (pps) with bombarding energies spanning from 10 up 100$A$~MeV. Such requirements highly constrain the number of available isotopes that can be studied via direct reactions in present facilities. Nonetheless, we are entering an era of next-generation radioactive beam facilities capable of producing sufficiently intense beams of exotic nuclei close to and beyond the drip lines. Some examples of these laboratories are the Facility for Rare Isotope Beams (FRIB) at the Michigan State University (USA)~\cite{York:2010lsa}, the Facility for Antiproton and Ion Research (FAIR) at GSI (Germany)~\cite{Eschke_2005}, HIE-ISOLDE at CERN (Switzerland)~\cite{BORGE2016408}, the Advanced Rare Isotope Laboratory (ARIEL) of TRIUMF (Canada)~\cite{Ball_2016}, or the future Rare isotope Accelerator complex for ON-line experiments (RAON, Korea) ~\cite{KIM2020408}.\\ 

The development of state-of-the-art instrumentation has progressed in parallel with the advancement of radioactive beam production. Pioneering direct-reaction experiments in inverse kinematics (transfer in particular) were performed using a telescope of solid state (namely silicon) detectors~\cite{WALLACE2007302,MUST2} and a thin composite target (e.g. CH$_2$). From the energy and angle of the light ejectile measured in the silicon detectors the kinematics of the reaction can be inferred with modest resolution (few hundred keV). This technique has been demonstrated in a plethora of successful experiments~\cite{Wimmer_2018}, but it has an important limitation: The energy is compressed due to the center-of-mass motion. As a result, the resolving power of these measurements is limited. This effect can be removed if the detection setup is placed inside a solenoid magnet. Such experimental devices are known as solenoidal spectrometers. The kinematic lines corresponding to the excited states of interest can be inferred by using the linear relation between the returning position of the spiraling particles along the solenoid axis and their energy measured in an array of silicon detectors deployed along the axis. The deleterious effects of kinematic compression are avoided in this approach, demonstrated for the first time using the HELIOS solenoidal spectrometer of the Argonne National Lab (ANL)~\cite{WUOSMAA20071290,LIGHTHALL201097}. The remarkable performance of this device has resulted in many outstanding transfer experiments with high-impact results~\cite{Back2010,Santiago2018}. Its success has also spurred the development of similar devices in different facilities, such as the Isolde Solenoidal Spectrometer (ISS) at CERN~\cite{Tang2020} and SOLARIS at FRIB~\cite{SOLARISWP}.\\

The outstanding resolution achieved by solenoidal spectrometers such as HELIOS depends strongly on the thickness of the target, which poses a limitation in terms of beam intensity. A typical thickness of a few hundreds of $\mu$g/cm$^{2}$ requires intensities of the order of 10$^{4-5}$ pps for single-nucleon transfer reactions. Higher intensities are needed for two-nucleon transfer reactions, as the cross sections are lower. Moreover, targets have contaminants (usually carbon) and are easily damaged under high-power irradiation. The acceptance of solenoidal spectrometers is also limited by the target thickness since particles with low energy are not able to escape from it. Naturally, the acceptance is also limited by the number of detectors and their geometry. In order to overcome these limitations, a different detection scheme can be adapted. For instance, the silicon array and target can be replaced by an Active Target Time Projection Chamber (TPC). In this case, a gas such as hydrogen, deuterium or helium, is used as a target and detector simultaneously. The gas is enclosed in a large volume with an intense electric field applied across. As charged particles ionize the gas, the ionization electrons
are drifted to a highly segmented pad plane where they are detected. This
is used to reconstruct the charged particles’ trajectories inside the volume~\cite{Ayyad2018}. This scheme brings many advantages over conventional setups: luminosity is increased by orders of magnitude, detection thresholds are lowered down to about 100~keV and the detection efficiency is increased close to 4$\pi$~\cite{BAZIN2020103790}. The magnetic field adds a robust observable, the magnetic rigidity, inferred from the track curvature and the magnetic field. The measurement of the energy loss together with the magnetic rigidity enables the identification of the particles. Moreover, the use of the magnetic field vastly extends the dynamic range of the detector. Because of such compelling capabilities, TPCs are gaining much interest for the study of radioactive nuclei far from stability~\cite{BECEIRONOVO2015124}. At present, only two TPCs operate inside a solenoidal field: the Active Target Time Projection Chamber (AT-TPC) of the Facility for Rare Isotope  (FRIB) of the Michigan State University (MSU)~\cite{BRADT201765} and the SpecMAT detector at ISOLDE~\cite{POLESHCHUK2021165765}. The first experiments with low-intensity radioactive beams using the AT-TPC have already demonstrated the advantages of this technique for direct reactions (scope of this work), resonance proton scattering~\cite{BRADT2018155} and reactions of astrophysical interest~\cite{RANDHAWA2020}.\\

Arguably, the most challenging aspect regarding the extraction of physical observables from TPC data is the reconstruction of kinematics from particle tracks. In general, this is a well established procedure in High Energy Physics (HEP) experiments where very efficient techniques have been developed over the course of many years~\cite{Fruhwirth429722}. However, the detector plus target scheme of Active Target TPCs leaves open challenges to address. Low-energy (on the range of 100 keV to several tens of MeV) reaction products slow down inside the TPC gas, tracing non-helical trajectories without an analytical description that complicates the extraction of relevant features from the three dimensional point cloud that represents the event recorded by the TPC~\cite{Ayyad2018}. In particular, the challenge lies in deducing the kinematic properties of the lowest energy particles that stop inside the volume and in inferring the reaction vertex and the track multiplicity. The latter has been successfully addressed using a non-parametric approach based on point triplets~\cite{DALITZ2019159}. However, an efficient and reliable method for determining the kinematics of particle tracks in solenoidal spectrometers working in Active Target mode is still required.\\

In this work, we describe the application of a linear quadratic estimator known as the Kalman filter~\cite{Kalman1960} for the fitting of particle tracks~\cite{FRUHWIRTH1987444} in Active Target TPCs. In particular, we benchmark the reconstruction of kinematics using data from two experiments performed with the AT-TPC: $^{10}\mathrm{Be}+d$ reaction at around 9$A$ MeV at ReA6 (FRIB) and the $^{14}\mathrm{C}+p$ reaction at 12$A$ MeV measured at ATLAS at Argonne National Laboratory using the AT-TPC inside the HELIOS magnet. Both experiments were performed with a low beam intensity of about 2000~pps for a total running time of 2 and 1 days for $^{10}$Be and $^{14}$C, respectively. Performing experiments with such a low intensity in a very short amount of time demonstrates how powerful Active Targets are for direct reactions in inverse kinematics with exotic beams. Comprehensive descriptions of the detector and the filtering method are presented in the first part of this manuscript. The performance of the track filtering is then evaluated using different reaction channels measured in these experiments.\\ 

\section{ Active target solenoidal spectrometers: Active Taget Time Projection Chamber (AT-TPC).} 
\label{sec:sec2}

The AT-TPC is a time projection chamber that simultaneously works as the reaction target. The observables inferred in the AT-TPC are the three dimensional tracks of the charged particles from a nuclear reaction that takes place between a radioactive beam and the gas used as the tracking medium. These tracks are a collection of points (hit pattern) deduced from the drift time of the ionization electrons and their two dimensional projection on the highly segmented pad plane of the detector. In particular, the AT-TPC features a cylindrical gas volume of 1~m length and 50~cm in diameter. The pad plane into segmented in 10,240 triangular pads. A more detailed description of the detector and its mode of operation can be found in Refs.~\cite{BRADT201765,Ayyad2018,AYYAD2018166,ZAMORA2021164899}.\\ 

\begin{figure} [htpb] 
  \centering
  \includegraphics [width=\linewidth] {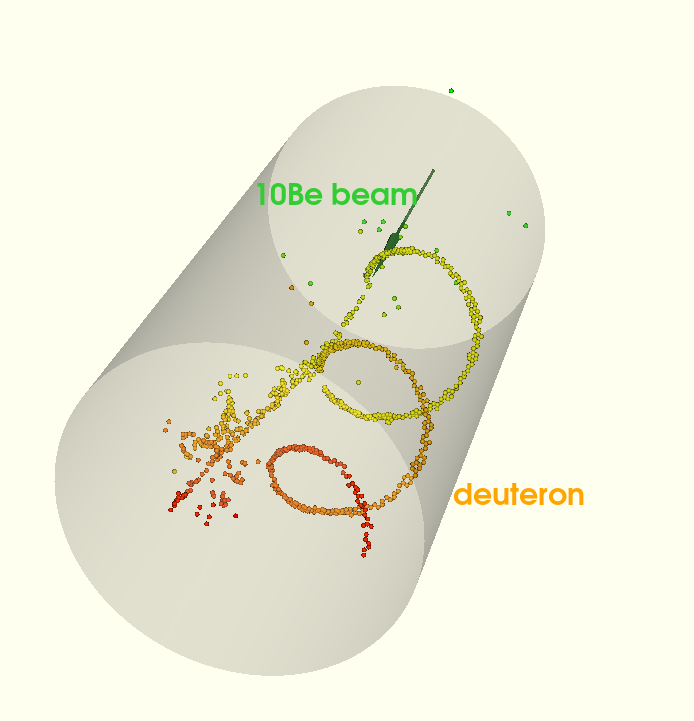}
  \caption{Hit pattern of a $^{10}\mathrm{Be}+d$ scattering event at 9$A$ MeV in a 3~T magnetic field in SOLARIS. Points are colored by the amplitude of the signal observed in the corresponding pad}
  \label{fig:fig_1}
\end{figure}
      
The AT-TPC is placed inside the 4-T solenoid of SOLARIS for the measurement of the magnetic rigidity of the particles. Typical magnetic fields of 2-4~T are used. Fig.~\ref{fig:fig_1} shows an example of the hit pattern of an event recorded with the AT-TPC: A $^{10}$Be beam of 10$A$ MeV (around 1000~pps of intensity) is injected along the AT-TPC beam axis, which is coincident with the solenoidal field axis. The detector is filled with pure D$_{2}$ gas at 600~torr. In the figure, the beam reacts with a deuteron which is scattered forward following a helical trajectory with ever decreasing radius. The AT-TPC offers excellent luminosity and angular coverage due to the large geometrical acceptance and the large dynamic energy range provided by the Multi-layer THick Gas Electron Multiplier (MTHGEM)~\cite{Cortesi2017}. The MTHGEM also enables the use of pure elemental gases as the target medium, making the AT-TPC the thickest pure target for direct reactions with around 10~mg/cm$^2$.\\

The AT-TPC and its smaller version, the prototype AT-TPC (pAT-TPC)~\cite{SUZUKI201239}, have been utilized in many successful experiments with radioactive beams to investigate the structure of exotic nuclei via resonance elastic scattering~\cite{BRADT201765}, reactions of astrophysical interest~\cite{RANDHAWA2020} and particle emission following $\beta$-decay~\cite{Ayyad2019}. Its scientific program encompasses diverse topics such as nucleus clustering, shell evolution, shape coexistence, giant resonances, and exotic decay modes. The extraction of relevant physics from these experiments requires a robust and efficient analysis method for curved trajectories.\\ 




\section{Low-energy particle track reconstruction with a Kalman Filter}

The Kalman filter is a recursive algorithm that provides the best estimate for a set of discrete experimental measurements affected by noise~\cite{Kalman1960}. The set of measurements represents the time-dependent state of a system or process with the best estimate provided by the minimization of the mean squared of the measurement errors. While the filter itself has many versions and applications, in this work we will just provide a brief outline of the mathematical formalism of the extended or nonlinear Kalman filter. More complete information on the application of Kalman filters in particle physics can be found in Refs.~\cite{FRUHWIRTH1987444,BILLOIR1985115,STAMPFER1994157,FRU2000}.\\

The idea behind Kalman filtering for particle tracks is rather powerful and simple. It requires a robust physics model, including experimental noise. Following the notation of Refs.~\cite{FRUHWIRTH1987444,HOPPNER2010518}, a set of measurements $m_{k}$ are used to find an optimum estimate $x_{k}$ of the true vector state $\hat{x_{k}}$. The evolution of the system can be described by the propagation of the track state for each measurement $k-1$ to state $k$, using the information from previous measurements up to $k-1$:

\begin{equation}
    x_k = f_{k-1}(x_{k-1}) + \omega_{k-1}
\end{equation}

where $\omega_{k-1}$ is the random (Gaussian) process noise due to the track propagation from $k$ to $k-1$, and $f_{k-1}$ is the track propagator described by the motion of the particle. In the traditional formulation, the energy loss of the particles is included as part of this process noise. This is not ideal in our case since the energy loss of heavy ions is much larger than that of minimum ionizing particles (MIPs) which are typically measured in high-energy
experiments. The proper way to account for the energy loss of the ions is by including it in the equation of motion, as we will discuss later. Since the state vector is not accessible directly, it is projected onto the $m_{k}$ space using a linear transformation function $h_{k}$ taking into account the Gaussian measurement noise $\epsilon_{k}$:

\begin{equation}
    m_k = h_{k}(x_{k}) + \epsilon_{k}
\end{equation}

In spite of $f$ and $h$ being non-linear functions, the evolution of the system's state vector is described by the associated linear transformations:

\begin{equation}
    x_k = F_{k-1}(x_{k-1}) + \omega_{k-1}
\end{equation}
\begin{equation}
    m_k = H_{k}(x_{k}) + \epsilon_{k}
\end{equation}

where $F$ and $H$ are first order Taylor expansions of their respective functions following the Extended Kalman Filter formalism~\cite{Daum2005}. For our particular case, $F$ follows the Runge-Kutta method to describe the movement of a charged particle inside the AT-TPC tracking medium. The use of a linear propagator required by the algorithm is ensured by using the first-order expansion of the Runge-Kutta propagator around the particle trajectory. Conveniently, the description of the state vector $x_{k}$ is chosen according to the topology of the TPC, where each hit in the space is defined in a local plane coordinate system, with two orthonormal vectors $u$ and $v$ with respect to the momentum of the particle $u'$ and $v'$~\cite{HOPPNER2010518,Rauch2021}:

\begin{equation}
\label{eq:eq_5}
    x_k = (q/p,u',v',u,v)^\mathrm{T}
\end{equation}

Based on this parameterization, $H$ transforms $u$ and $v$ into the $m_{k}$ system.\\

The fitting process is performed in three steps: Prediction, filtering and smoothing. During the prediction state, the last known state and its covariance $C$ are extrapolated to the present state based on all previous measurements up to $k-1$ (indicated by the second index on the left side of the equations):

\begin{equation}
     x_{k|k-1} = F_{k-1}(x_{k-1}) + \omega_{k-1}
\end{equation}
\begin{equation}
     C_{k|k-1} = F_{k-1}C_{k-1}F^T_{k-1} + N_{k-1}
\end{equation}

where $N_{k-1}$ is the covariance matrix of the propagation noise. The predicted state at $k$ is updated (filtered) based on the measurement $m_{k}$ and the weight of the residual $K$, also called Kalman gain:

\begin{equation}
     x_{k} = x_{k|k-1} + K_{k}( m_k - H_k x_{k|k-1} )
\end{equation}
\begin{equation}
     C_{k} = (I - K_k H_k) C_{k|k-1} 
\end{equation}

The Kalman gain gives a quantitative estimate of how much the estimate has to change based on the measurement. Lastly, the filtered track is also fitted in backward direction using the information from previous measurements ($n$ index) to provide a smoothed track:

\begin{equation}
  x_{k|n} = x_{k} + A_{k} ( x_{k+1|n} - x_{k+1|k})
\end{equation}
\begin{equation}
 C_{k|n} = C_{k} + A_{k} (C_{k+1|n} - C_{k+1|k}) A^T_{k-1} 
\end{equation}

where $A_{k}$ stands for the smoother gain matrix: 

\begin{equation}
A_{k} = C_{k} F^T_{k} C_{k+1|k}
\end{equation}

To summarize the procedure for our case of interest: For the Kalman fitting process described above, the hit pattern is used as measurement points together with a representation of the particle track characterized by the motion of a low-energy charged particle under the effect of the solenoid field in a gaseous medium. Points are clusterized (cluster hits) along the particle trajectory to define virtual detection planes following Eq.~\ref{eq:eq_5}. Multiple scattering and straggling account for the process noise in the track propagation. The definition of the covariance matrix is based on the position uncertainty determined by the diffusion coefficients, pad plane granularity and the sampling rate~\cite{Shi2017}. In order to account for these effects more typical of low-energy nuclear physics experiments, we have modified the package \textsc{genfit}~\cite{HOPPNER2010518,Rauch2021} to fit AT-TPC data and investigated its performance. The code was integrated into the \textsc{ATTPCROOTv2} analysis framework~\cite{Ayyad_2017,ATTPCROOT_git,GIRAUD2023168213}.\\

\section{Event and kinematics reconstruction}

\begin{figure} [htpb] 
  \centering
  \includegraphics [scale=0.45] {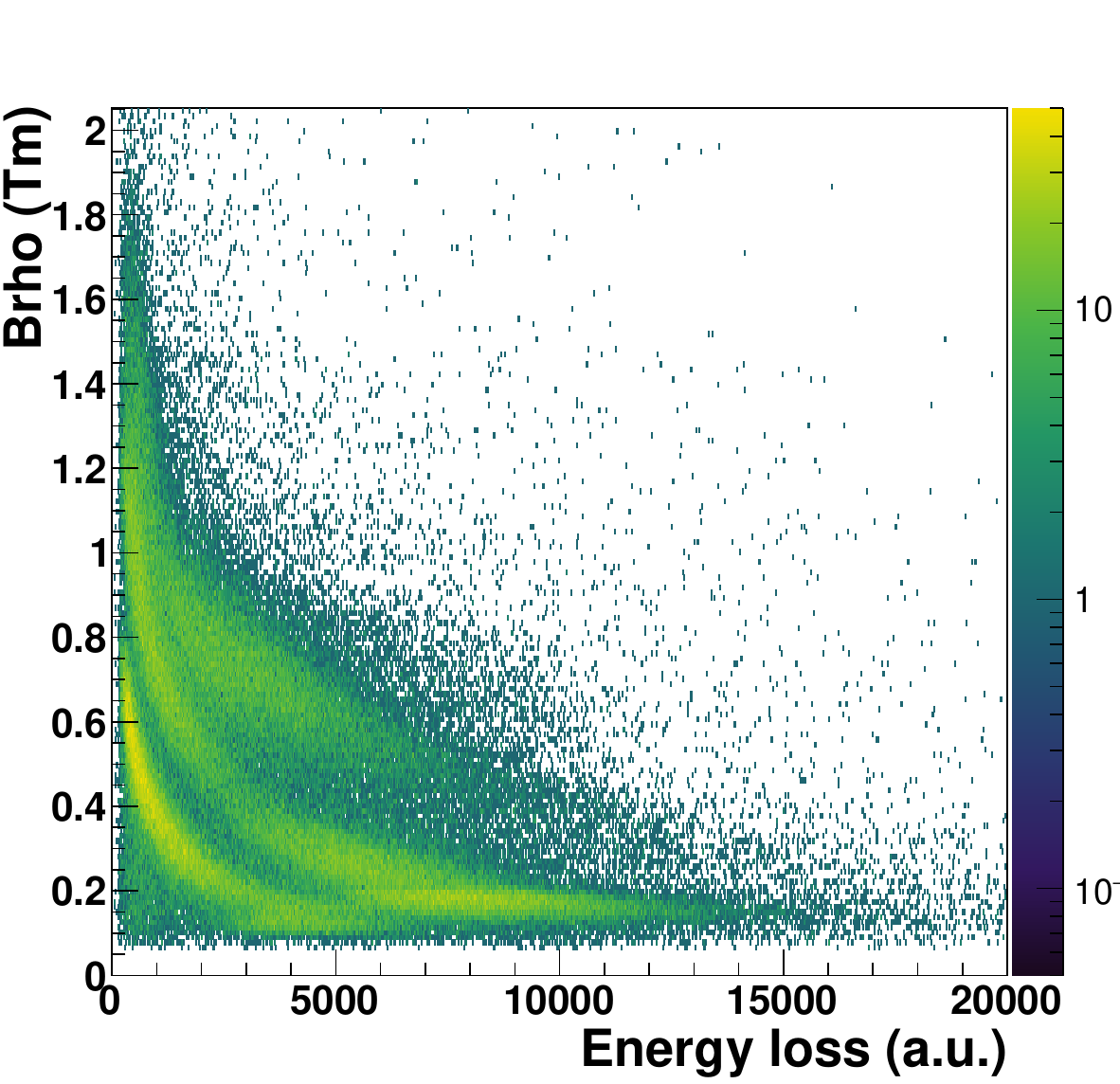}
  \caption{B$\rho$ as a function of the energy loss inferred from the track geometry. The upper, middle and lower regions correspond to beryllium and $\alpha$ particles, deuterons and protons, respectively.}
  \label{fig:fig_2}
\end{figure}

The reconstruction of the hit pattern shown in Fig.~\ref{fig:fig_1} is done using the 512-samples trace recorded by each pixel of the Micromegas sensor~\cite{GIOMATARIS199629} upon the arrival of the electrons. From the shape of these traces the $z$ position and energy loss are inferred from their time at the peak and their amplitude. This simple approach is reasonable for scattered target nuclei because the shape of the trace is dominated by the shaping time of the electronics. Once the hit pattern is constructed, single tracks corresponding to different reaction products are identified using a dedicated clustering algorithm designed for the characteristic tracks of low-energy particles, as explained in Ref.~\cite{DALITZ2019159}. At this point, each track can be pre-analyzed independently to extract the initial conditions for the fitting. Hits are clusterized in charge clusters of a certain radius and separation along the trajectory. The position of these new charge clusters is calculated using the center of gravity of the collection of hits that belong to each of them. The purpose of this process is two fold: smoothing the trajectory and preparing the track for the filtering process. Particle identification can be performed for each track from the respective magnetic rigidity and energy loss. The rigidity (B$\rho$) is calculated from the radius of curvature of the track inferred from its geometry~\cite{AYYAD2018166}. The energy loss is inferred from the charge collected over the first 30$\%$ of the track, averaged by the number of charge clusters.  The identification plot for the $^{10}\mathrm{Be}+d$ reaction, shown in Fig.~\ref{fig:fig_2}, clearly shows three regions corresponding to protons, deuterons and overlapping $\alpha$ particles and beryllium isotopes.\\

Once scattered deuterons are selected in the particle identification plot, their energy and scattering angle are inferred from the B$\rho$ and the angle that the track forms with the $Z$ (beam/solenoid) axis, as explained before. The result of this geometrical estimate serves as the starting point for the Kalman filtering process. Cluster hits are used as measurement points with an associated covariance matrix defined by uncertainties in the hit-position determination. Each point acts as a virtual plane to perform the propagation of the track based on the chosen model. The state vector $x_{k}$ is parameterized according to the initial vertex and momentum deduced from the rigidity. One of the main limitations of ~\textsc{genfit} regarding our scope is the treatment of the energy loss. Due to the complex charge-exchange interplay of low-energy heavy ions when traversing a gas, the default Bethe-Bloch energy loss formalism used in the code is not valid anymore. The evaluation of the energy loss between hit clusters was modified to include a parameterization based on the ~\textsc{SRIM} code~\cite{ZIEGLER20101818}. Even though \textsc{genfit} was developed as a generic tool for tracking, it treats the energy loss as part of the process noise because it is generally used, to the best of our knowledge, to reconstruct the minimum ionizing particles in TPCs. The effect of the energy loss in the AT-TPC is large enough to be included as part of the equation of motion, as particles are continuously slowed down inside the detector describing complicated trajectories. This feature may pose a limitation in terms of resolution and performance. At the prediction stage, the hit clusters are accepted or rejected for filtering based on the initial trajectory. Once the initial conditions are defined, the fitting is realized on a recursive basis via filtering and smoothing. The result from the fit is a representation of the track from which the best estimate of the momentum for a given particle is inferred. Moreover, this track representation can be used to find the reaction vertex by extrapolating the first hit of the track to the pad plane origin.\\

\begin{figure} [htpb] 
  \centering
  \includegraphics [scale=0.33] {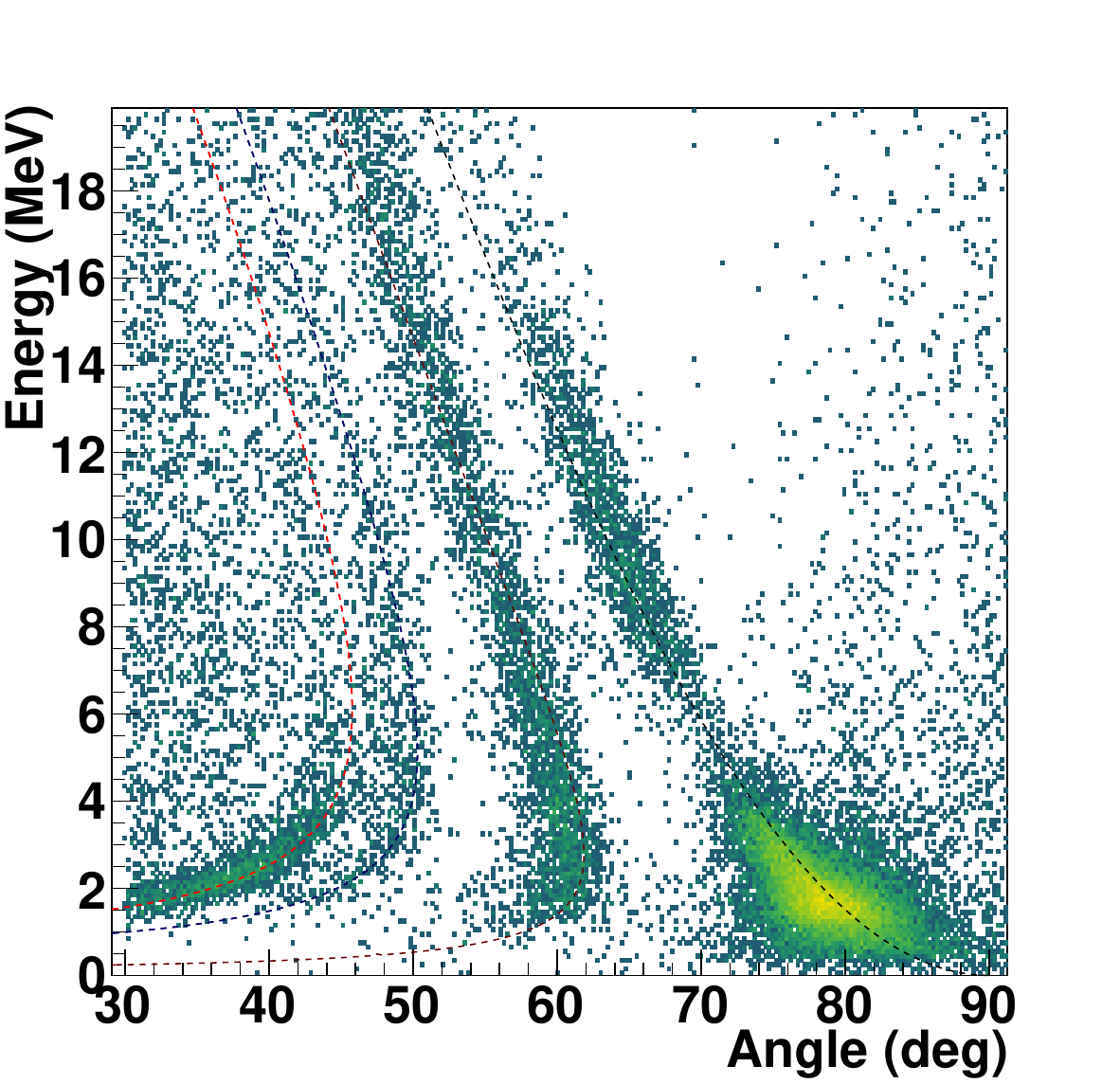}
  \includegraphics [scale=0.33] {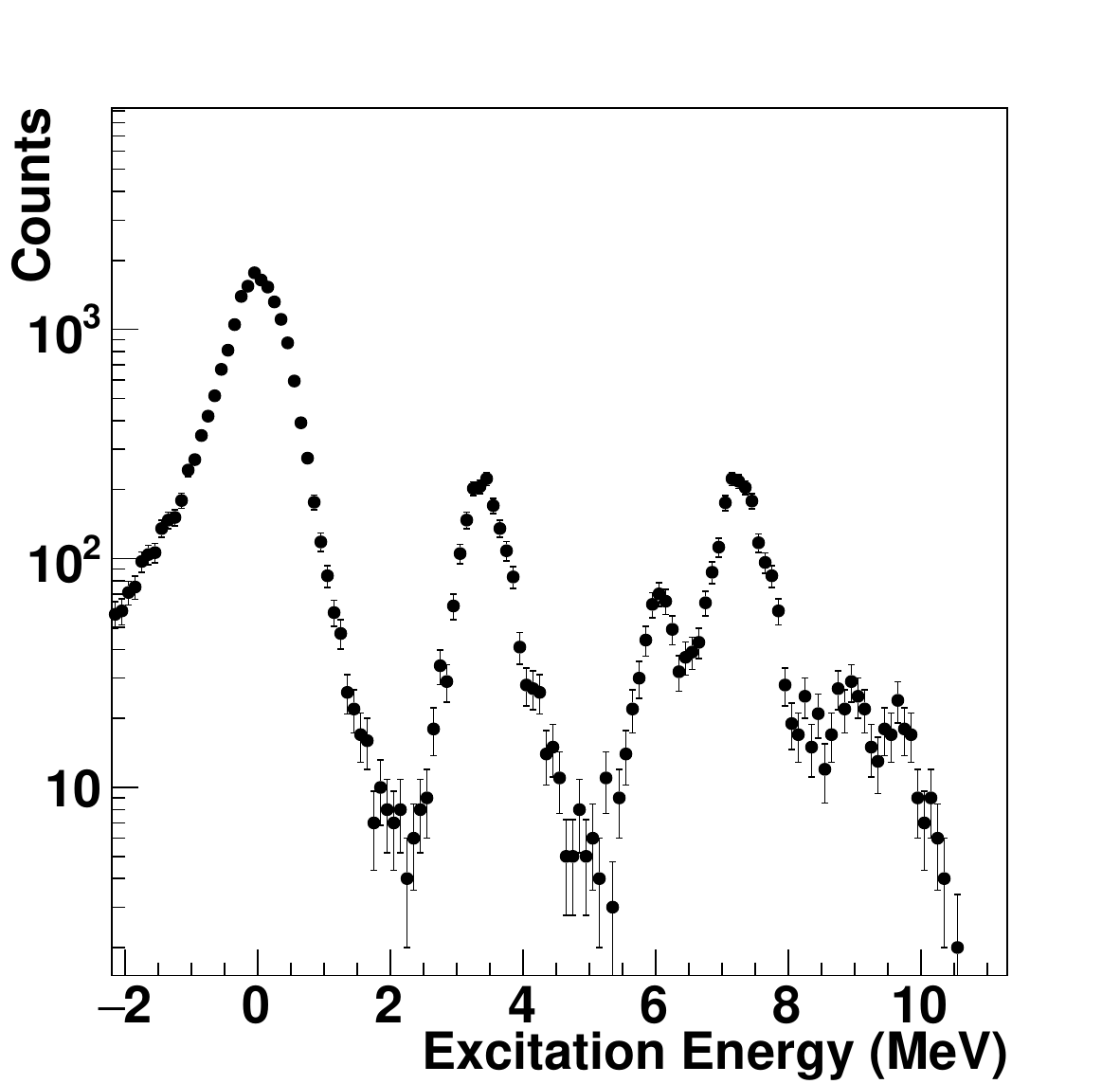}
  
  \caption{Left panel: $^{10}$Be+d kinematics reconstructed from the track geometry for reactions along the entire AT-TPC length. Right panel: Excitation energy distribution of $^{10}$Be from inelastic scattering .}
  \label{fig:fig_3}
\end{figure}

Results from the filter are shown in Fig.~\ref{fig:fig_3}. The kinematics plot, shown in the left panel of Fig.~\ref{fig:fig_3}, features the  characteristic kinematic lines of several $^{10}$Be states: ground state (0$_{1}^{+}$), first excited state 3.368~MeV (2$_{1}^{+}$), a multiplet of three peaks at around 6~MeV and another state at around  7.2~MeV. The angular distribution of the latter does not fit with the spin-parity assignment of the two known states around that region (7.371~MeV with 3$_{1}^{-}$ and 7.542~MeV 2$_{3}^{+}$). The properties of that state will be discussed in a separate publication. The effect of the fitting is two-fold: it improves the resolution inferred from the track geometry to 350~keV (standard deviation) and the accuracy (about 5~keV) for the entire length of the detector. It is worth noting that the tracking efficiency decreases with the scattering angle, increasing the energy detection threshold (our limit is about 0.5~MeV). The main reason for this efficiency loss is the limited performance of \textsc{genfit} when reconstructing short tracks with scattering angles around 90$^{\circ}$, as the code was not designed for these cases. Below 40$^{\circ}$, there is a region of events that corresponds to non-resonant reactions above particle emission threshold but also to misidentified deuterons.\\

\begin{figure} [htpb]
  \centering
  \includegraphics [scale=0.33] {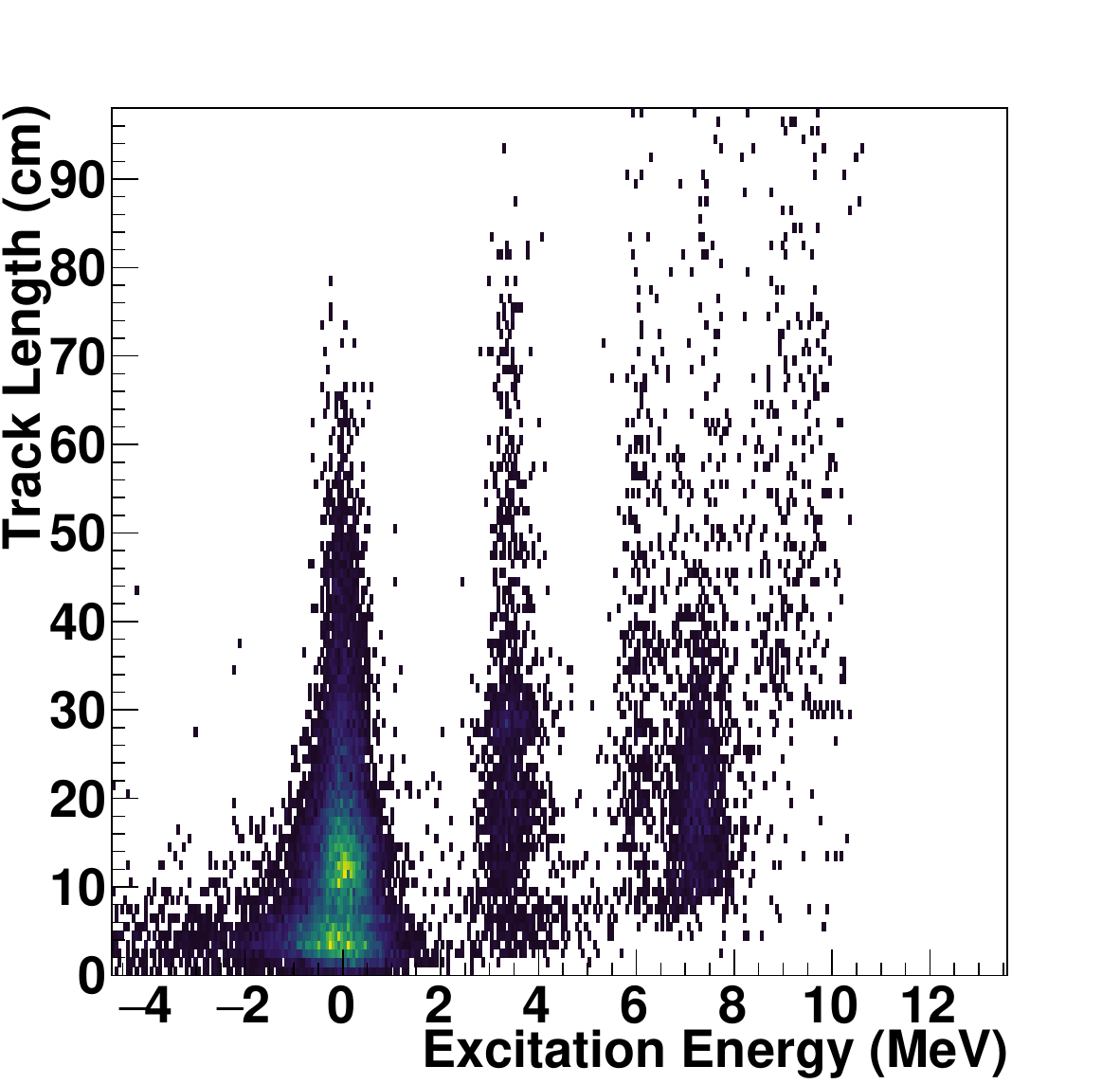}
  \includegraphics [scale=0.33] {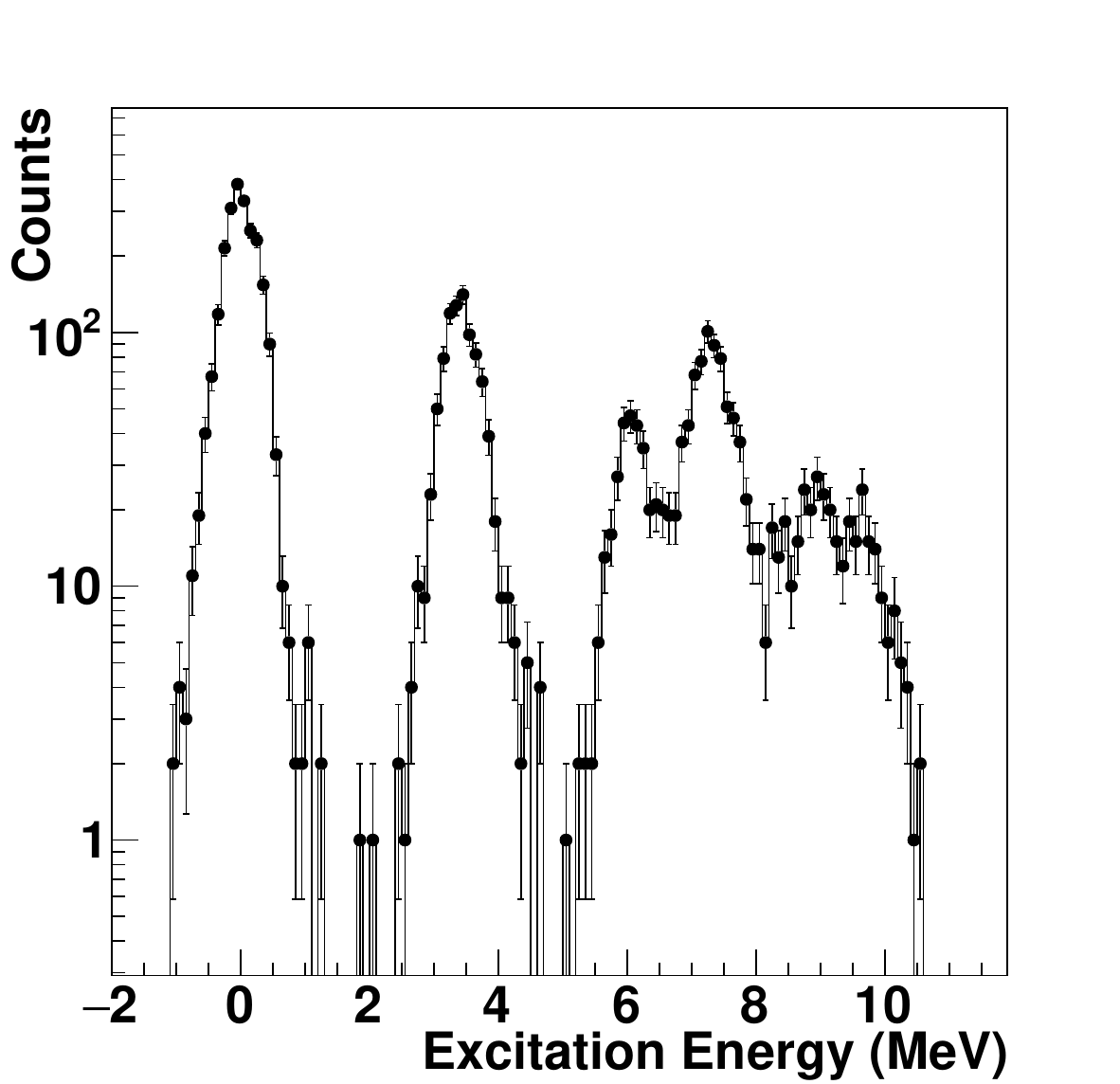}
  \caption{Left panel: Track length as a function of the excitation energy. Right panel: Excitation energy of $^{10}$Be after selecting tracks of more than 20~cm length. }
  \label{fig:fig_4}
\end{figure}

The energy resolution depends on several fit parameters, among many others related to the detector configuration. The length of the fitted track is one of the most critical parameters. Designing an experiment with the AT-TPC necessitates several trade-offs depending on the goal of the experiment. For example, lowering the pressure will decrease the statistics, but increase the length of the tracks thereby increasing the energy resolution. The selection of such parameters is critical for experiments aiming to measure reactions at very forward center-of-mass angles. The measurement of the $^{10}\mathrm{Be}+d$ reaction was not optimized for detecting and measuring low-energy tracks, but the resolution can be improved by studying the dependence of the excitation energy on the length of the track and the energy of the deuterons (right panel). Figure~\ref{fig:fig_4} shows the excitation energy as a function of the track length (left panel) and its projection after selecting tracks of more than 20~cm length and deuterons below 10~MeV kinetic energy. It is important to point out that the track length represents the useful part of the track that was used for the fitting procedure as \textsc{genfit} rejects clusters where the propagation between virtual planes failed. This usually happens when the cluster contains many noisy points that were assigned as inliers by the clustering algorithm due to their proximity to the track. In any case, one can see a clear correlation between the track length (proportional to the number of fitted clusters) and the resolution. As expected, more information results in a better determination of the excitation energy, reducing the resolution down to 240~keV (standard deviation).  Therefore, depending on the experiment, the gas parameters can be adjusted to either provide the best possible resolution (i.e. decreasing pressure for very-forward angle measurements) or the best luminosity (e.g. experiments with broad resonances).

Another critical aspect for the reconstruction of tracks are the secondary effects that impact the propagation of the scattered particles in the gas. Although \textsc{genfit} includes multiple scattering and straggling effects as gaussian noise in the filtering process, the models are more adequate for MIPs with much larger kinetic energy than our ions. As a consequence, the update of the state vector and the covariance matrices may be biased due to the underestimation of such effects. Ideally, one would expect to obtain a much better resolution if realistic effects are included in the estimation of the track parameters. However, the complexity of \textsc{genfit} source code renders its modification complicated. Instead, we have studied the impacts of secondary effects applying the same reconstruction procedure to the $^{14}\mathrm{C}+p$ reaction, where we used 300~Torr of pure H$_2$ gas and a magnetic field of 2.85~T using the HELIOS magnet.\\

\begin{figure} [htpb] 
  \centering
  \includegraphics [scale=0.33] {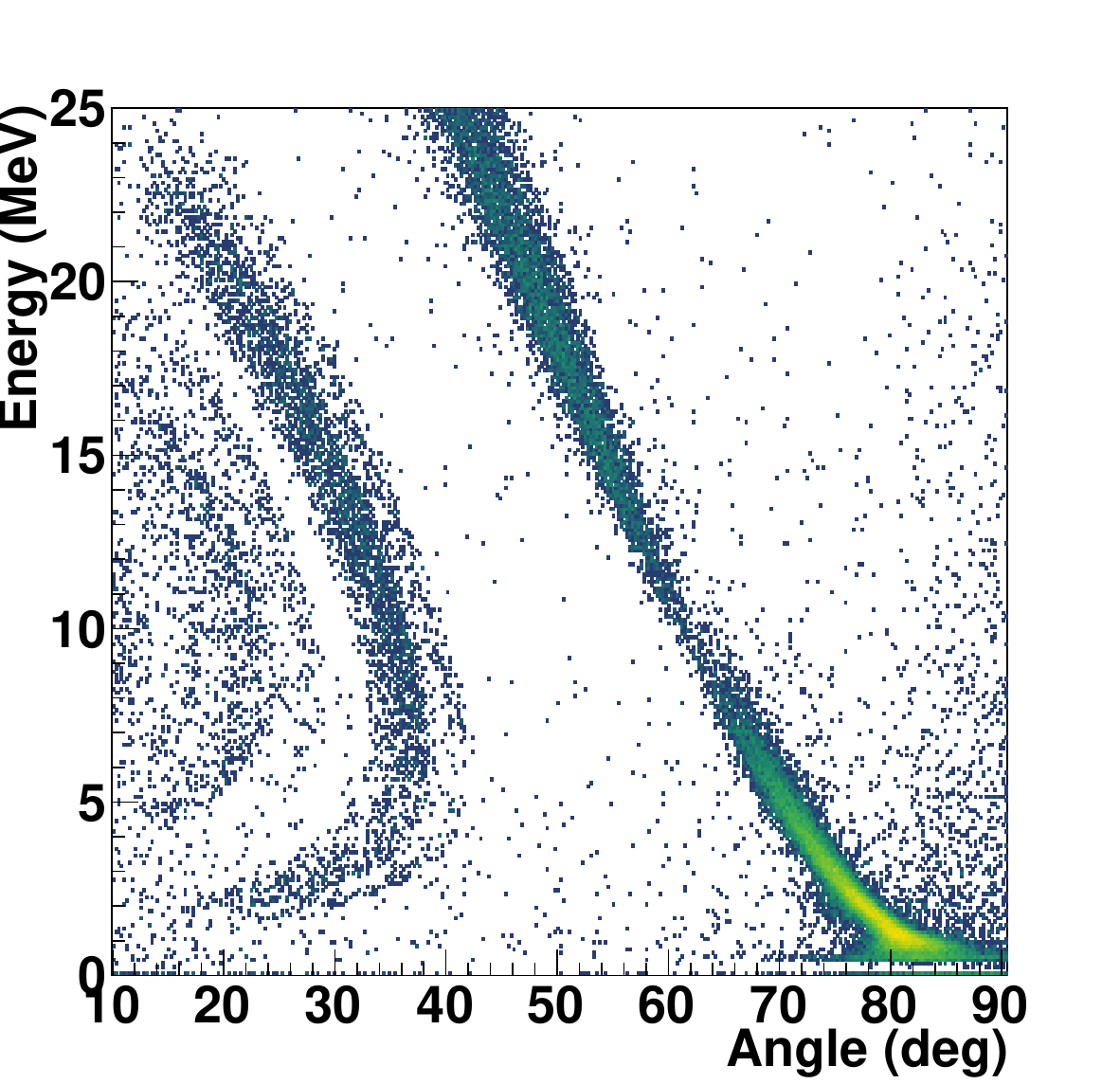}
  \includegraphics [scale=0.33] {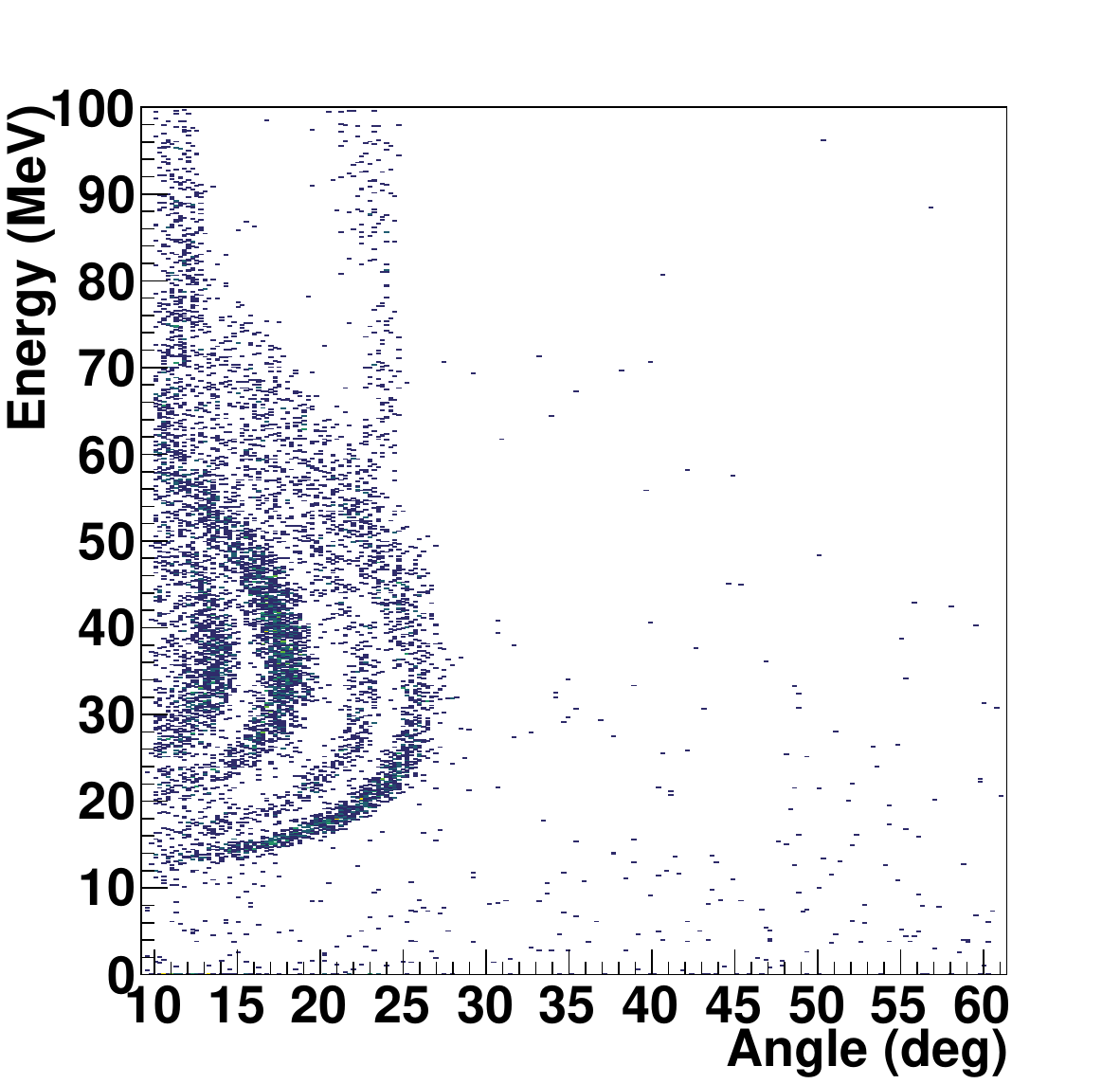}
  \includegraphics [scale=0.33] {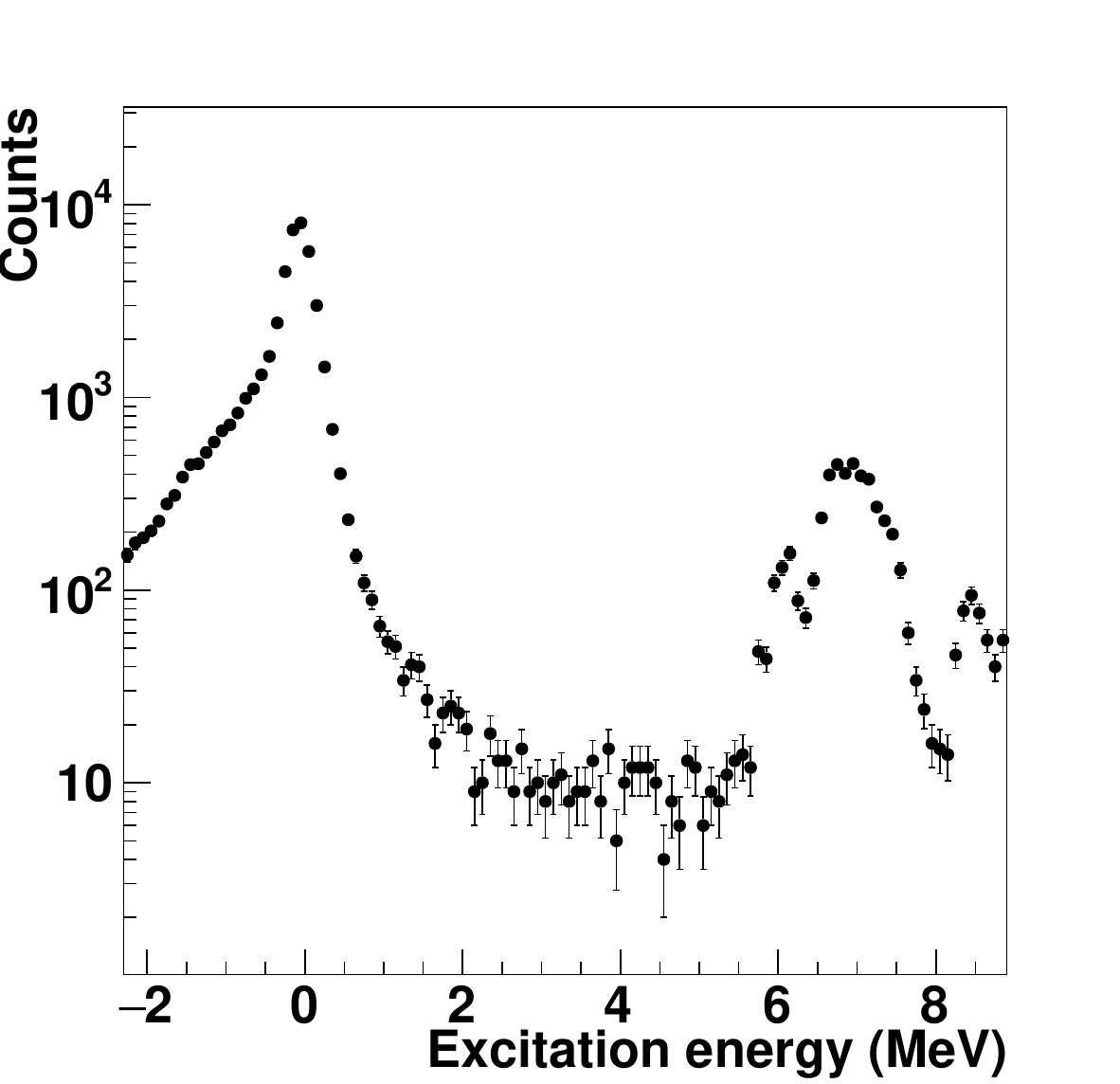}
  \caption{Left upper panel: Kinematics for the $^{14}$C(p,p) and $^{14}$C(p,p') reaction. Right upper: Same as left but for the $^{14}\mathrm{C}(p,d){}^{13}\mathrm{C}$, $^{14}\mathrm{C}(p,\alpha){}^{11}\mathrm{B}$ and the $^{14}$C(p,t)$^{12}$C reactions, measured simultaneously. Lower panel: Excitation energy of $^{14}$C (see text for details).}
  \label{fig:fig_14C}
\end{figure}

The results shown in Fig.~\ref{fig:fig_14C} for scattering (left upper panel) clearly indicate, at first glance, that the performance achieved for proton tracking is superior in terms of resolution. Moreover, the absence of characteristic background due to deuteron breakup observed in the $^{10}\mathrm{Be}+d$ (Fig.~\ref{fig:fig_3}) makes the reconstruction cleaner. Overall, we have reconstructed the $^{14}$C ground and several excited states: 6.093 (1$^{-}$) and 8.317 (2$^{+}$) (see lower panel of Fig.~\ref{fig:fig_3}). The group of unresolved states are assigned, within our uncertainty, to 6.589 (0$^{+}$),  6.728~MeV (3$^{-}$), 7.012 (2$^{+}$) and 7.341~MeV (2$^{-}$), based on the National Nuclear Data Center database. This assignment is consistent with the only $^{14}$C+p reaction performed to date that covers up to 9~MeV in excitation energy~\cite{LOZOWSKI198954}. The best resolution achieved for this spectrum is about 145~keV (standard deviation), a very competitive value for reactions in inverse kinematics taking into account that these data were acquired at a bombarding intensity of 2000~pps. Simultaneously to the inelastic scattering of $^{14}$C on protons, we have also measured the neutron removal $^{14}\mathrm{C}(p,d){}^{13}\mathrm{C}$, two-neutron removal $^{14}$C(p,t)$^{12}$C and the $^{14}\mathrm{C}(p,\alpha){}^{11}\mathrm{B}$ transfer reactions, as can be seen in the kinematics plot in the right upper panel of Fig.~\ref{fig:fig_14C}. These reaction channels were identified through the B$\rho$-Energy loss correlation, as shown in Fig.~\ref{fig:fig_2}, but for the $^{14}\mathrm{C}+p$ reaction. Since deuterons and $\alpha$ particles have the same rigidity (or mass-to-charge ratio), the apparent kinematic lines overlap within the same energy range as the $B\rho$ calculation was done assuming protons. The interesting point about the comparison between inelastic scattering and transfer is the excitation energy resolution. For the $(p,d)$ reaction, the excitation energy resolution is worse than in the $(p,p')$ case (about 220~keV). This resolution is comparable to the one obtained for the $^{10}\mathrm{Be}+d$, but at higher gas pressure. Therefore, this result suggests that the degradation of excitation energy has strong dependence on the straggling caused by a higher-mass particle, within the range of gas pressures studied in this work.\\

\section{Simulations and angular distributions}

Due to the complexity of the Kalman Filter package we are using, it is difficult to evaluate the absolute tracking efficiency. The high density of points in the hit pattern poses a problem for fitting convergence due to presence of outliers in some of the tracks. In addition, \textsc{genfit} was not designed for tracking particles that stop in the detection medium and therefore, the loss of efficiency at lower energy is more pronounced in the case of deuteron scattering. Following a rather pragmatic procedure, we corrected the angular distributions for the angle- and energy-dependent tracking efficiency provided by \textsc{genfit}. The correction was inferred from Monte Carlo simulations performed with \textsc{ATTPCROOTv2}. We simulated the $^{10}\mathrm{Be}+d$ reaction populating the ground state and the first excited state of $^{10}$Be using a flat angular distribution covering the same angular domain as the experimental data. The first part of the simulation involves the transport of charged particles in the detector, generating a collection of spatial points with an associated energy loss. At each point, a cloud of ionization electrons is generated based on the gas ionization potential. Each individual electron is transported to the pad plane where it is multiplied by avalanche and assigned to a certain pad based on its drift velocity and its lateral and longitudinal diffusion coefficients (ignoring the charge spread of the avalanche). The number of electrons collected in each pad along with their time of arrival are used to generate a pulse based on the electronics settings, namely the shaping time and the gain~\cite{POLLACCO201881,Obertelli2014}. The final pulse is the convolution of the current collected on each pad with the response function of the electronics. At this point, the simulated track reconstruction procedure follows the same steps as the experimental one and therefore the same reconstruction algorithms can be used.\\         

\begin{figure} [htpb]
  \centering
  \includegraphics [scale=0.33] {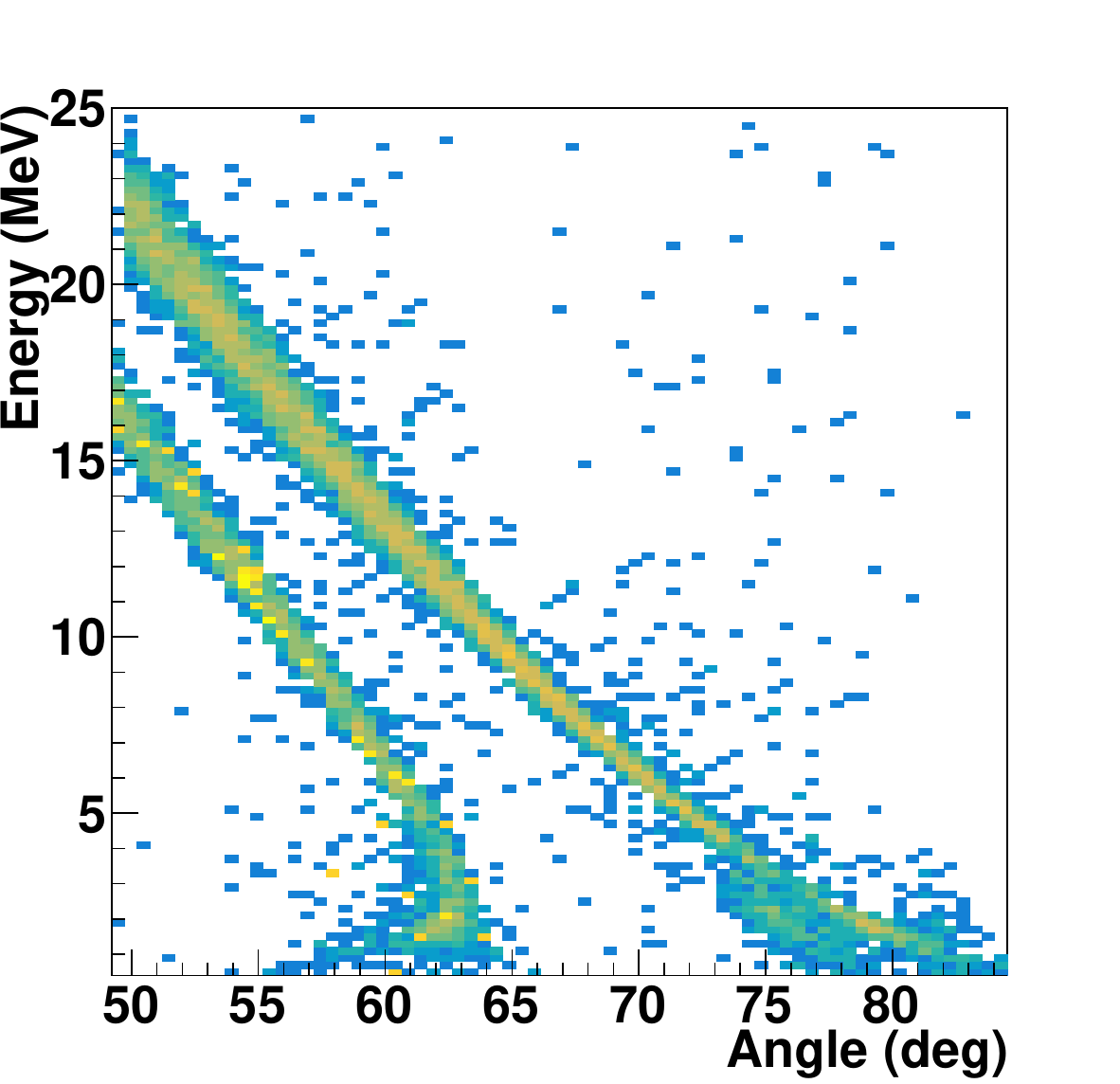}
  \includegraphics [scale=0.33] {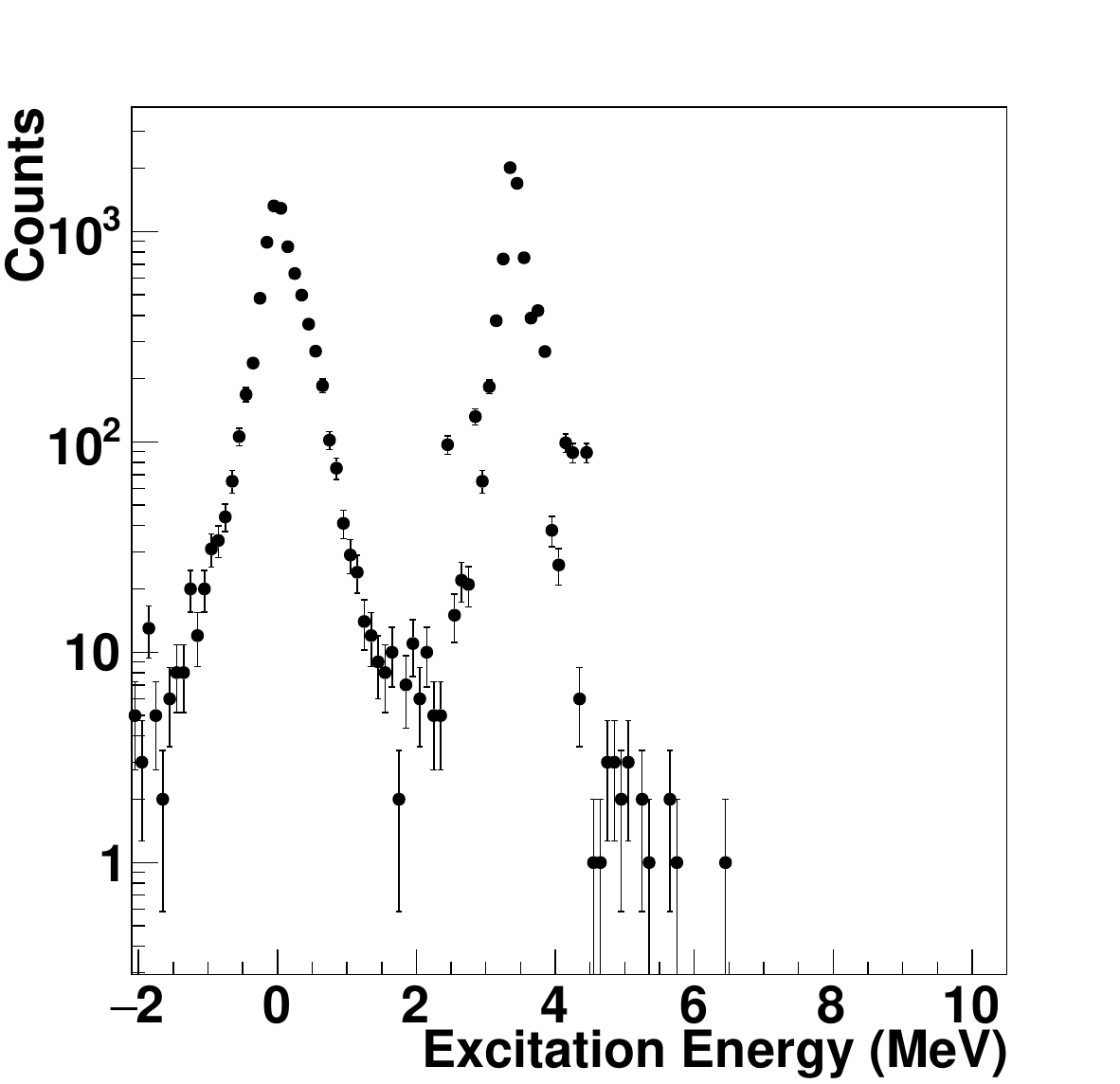}
  \caption{Left panel: Kinematics of the simulated $^{10}\mathrm{Be}+d$ reaction showing the ground and the first excited state. Right panel: Excitation energy of $^{10}$Be inferred from the simulation. Only two well-known states were used on this simulation.}
  \label{fig:fig_6}
\end{figure}

Figure \ref{fig:fig_6} shows the simulated kinematics and the excitation energy of the $^{10}\mathrm{Be}+d$ reaction. The excitation energy resolution amounts to 265~keV and 190~keV (standard deviation) for the ground and first excited state, respectively. In general, the simulation reproduces the experimental data rather well, in particular, the energy resolution, although some physical effects may be underestimated. The broadening of the kinematics at low kinetic energy is consistent with the experimental observation in Fig.~\ref{fig:fig_3}. The degradation of the resolution at around 90$^{\circ}$ seems to be underestimated  in the simulation compared to the experimental data. The efficiency correction function was inferred from the ratio of reconstructed and simulated events in the center of mass (CM) frame.\\ 

\begin{figure} 
  \centering
  \includegraphics [scale=0.33] {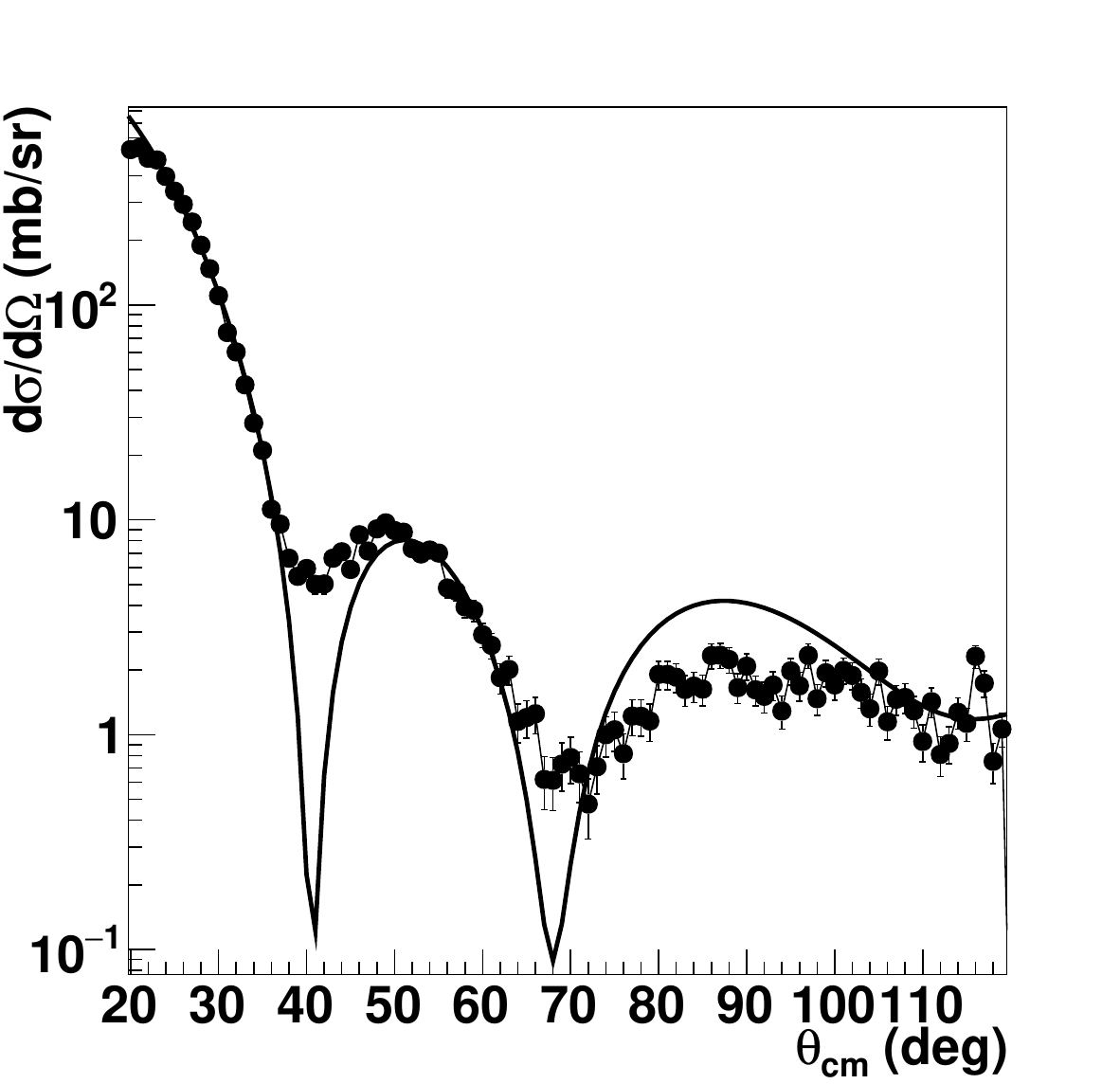}
  \includegraphics [scale=0.33] {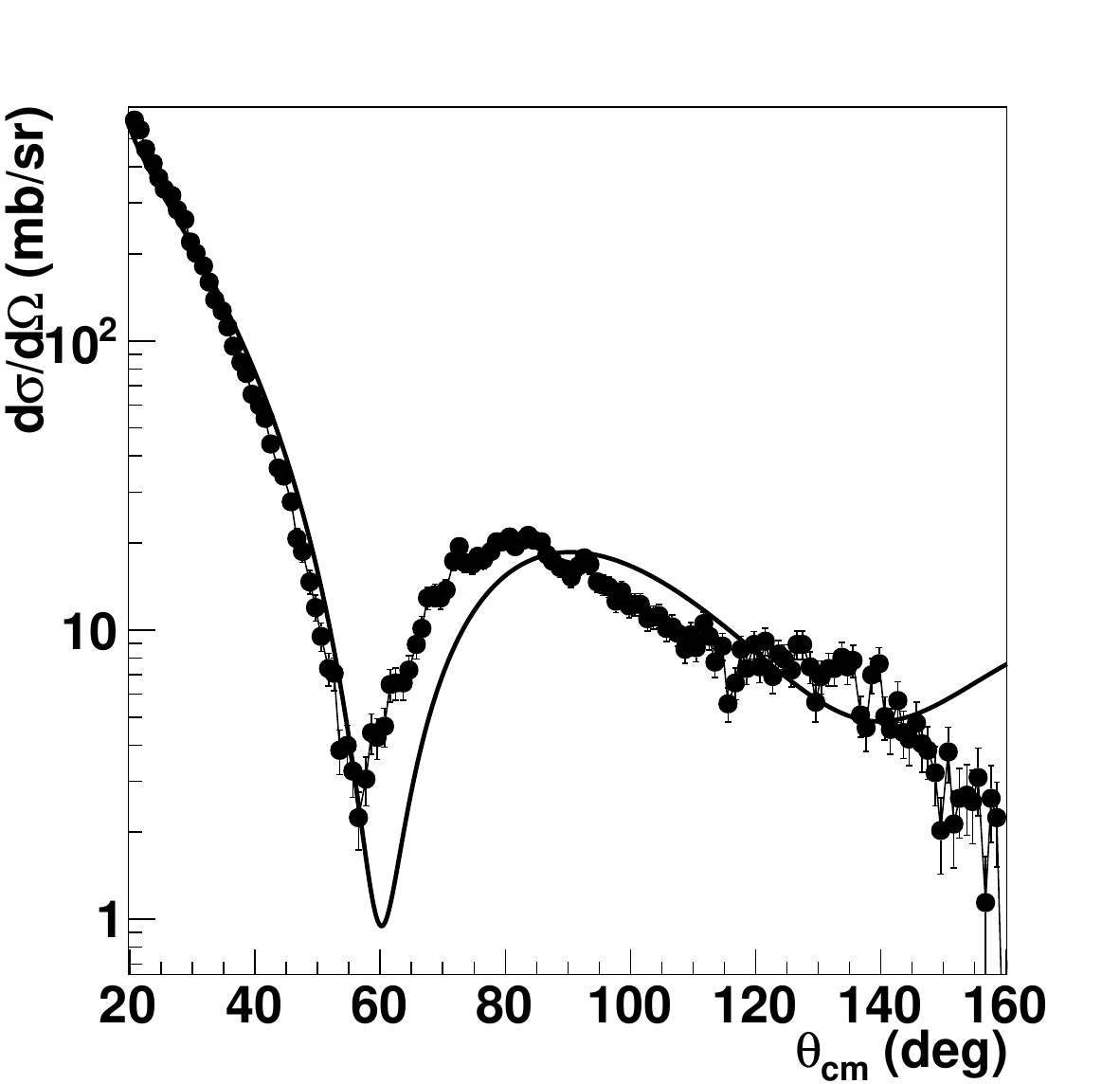}
  \caption{ Left panel: Angular distribution of the $^{10}$Be+d elastic scattering corrected by efficiency. The solid circles and the solid line are the experimental data and the DWBA theoretical distribution. Right panel: Same as left panel but for the $^{14}$C+p elastic scattering.  }
  \label{fig:fig_7}
\end{figure}

The correction function was used to correct the experimental angular distribution of the $^{10}\mathrm{Be}+d$ elastic scattering, shown as solid circles in the left panel of Fig.~\ref{fig:fig_7}. The experimental angular distribution was normalized by the target thickness and the beam intensity, and corrected bin by bin (1$^{\circ}$ size) with the results from the simulation. Below 20$^{\circ}$ CM (not shown) the efficiency falls rapidly because of the combined effect of the tracking and detection acceptance. The rest of the distribution has a rather good agreement with a simple optical model fit performed with \textsc{ptolemy}~\cite{Macfarlane1978} (solid line). On the right panel of Fig.~\ref{fig:fig_7}, we show the angular distribution for the $^{14}\mathrm{C}+p$ elastic scattering also compared to a DWBA calculation. In this case, the normalization of the data was based on the calculation. The efficiency correction is more critical at forward angles where the efficiency drops because the performance of the reconstruction package is very sensitive to the length of low energy particles tracks. As expected, this effect is more pronounced for deuteron scattering than proton elastic scattering where the correction is basically a scaling factor over the entire angular domain covered by the detector. The most remarkable aspect regarding the angular distributions is the large range covered which enables a precise adjustment of the associated optical potentials and the determination of physical quantities, such as matter deformation lengths, with much better precision~\cite{Chen2022}. 

\section{Conclusions and outlook}

In this work we have presented a tracking algorithm for the kinematics reconstruction of reactions in inverse kinematics using a solenoidal spectrometer in active target time projection chamber mode. The algorithm is based on the well-known extended Kalman filter implemented in the \textsc{genfit} package but adapted for low energy particles within a quite broad phase space. In particular, we have modified the code to use a more realistic description of the energy loss for ions traversing a gaseous material, which is rather critical at low energy. We have implemented \textsc{genfit} as an additional sequential task inside our \textsc{ATTPCROOTv2} analysis framework. The performance of the entire reconstruction procedure, and particularly the track reconstruction algorithm, was evaluated using data acquired with the AT-TPC in two experiments in inverse kinematics with radioactive beams on proton and deuteron targets.\\

We have successfully identified several reaction channels (scattering and transfer) through the identification of charged particles produced in these reactions. By fitting the particle tracks, it was possible to extract the reaction kinematics and the excitation energy spectrum. Overall, we have obtained an excitation energy resolution ranging from 150 to 350~keV (standard deviation), depending on the target. This method, together with the outstanding performance of the detector, allowed for the determination of a broad angular distribution covering almost 100$^{\circ}$ in CM. We also simulated the scattering reaction to extract the correction factors for the angular distributions to account for the efficiency loss.\\ 

The results obtained in this study clearly demonstrated that the Kalman Filter is a rather powerful tracking algorithm for kinematics reconstruction in solenoidal spectrometers. Developing a dedicated Kalman Filter for low energy particles with complicated trajectories requires adding several ingredients including a more robust treatment of the motion of low-energy particles in the medium and a realistic description of secondary scattering and straggling effects. We expect to improve the energy resolution if these effects are taken into account. Such a dedicated algorithm is currently under development by our collaboration.\\   

\section{Acknowledgements}

This material is based upon work supported by NSF’s National Superconducting Cyclotron Laboratory, which is a major facility fully funded by the National Science Foundation under award PHY-1565546. This material is also based upon work supported by the U.S. Department of Energy, Office of Science, Office of Nuclear Physics and used resources of the Facility for Rare Isotope Beams (FRIB), which is a U.S. DOE Office of Science User Facility under Award No. DE-SC0000661 SOLARIS is funded by the DOE Office of Science under the FRIB Cooperative Agreement DE-SC0000661. This material is based upon work supported by the U.S. Department of Energy, Office of Science, Office of Nuclear Physics, under Contract No. DE-AC02- 06CH11357 (Argonne). This research used resources of ANL’s ATLAS facility, which is a DOE Office of Science User Facility. This work has received financial support from Xunta de Galicia (CIGUS Network of Research Centers). Y. A. acknowledges the support by the Spanish Ministerio de Economía y Competitividad through the Programmes “Ramón y Cajal” with the Grant No. RYC2019-028438-I.


\appendix

 \bibliographystyle{elsarticle-num} 
 \bibliography{cas-refs}

\begin{thebibliography}{10}
\expandafter\ifx\csname url\endcsname\relax
  \def\url#1{\texttt{#1}}\fi
\expandafter\ifx\csname urlprefix\endcsname\relax\def\urlprefix{URL }\fi
\expandafter\ifx\csname href\endcsname\relax
  \def\href#1#2{#2} \def\path#1{#1}\fi

\bibitem{osti_1296778}
Reaching for the horizon: The 2015 long range plan for nuclear science (9
  2015).

\bibitem{Shapira1985}
D.~Shapira, Nuclear reaction studies using inverse kinematics, Tech. rep.,
  United States, cONF-8509176-- (1985).

\bibitem{Wimmer_2018}
K.~Wimmer, Nucleon transfer reactions with radioactive beams, Journal of
  Physics G: Nuclear and Particle Physics 45~(3) (2018) 033002.
\newblock \href {https://doi.org/10.1088/1361-6471/aaa2bf}
  {\path{doi:10.1088/1361-6471/aaa2bf}}.

\bibitem{Catford2014}
W.~N. Catford, What Can We Learn from Transfer, and How Is Best to Do It?,
  Springer Berlin Heidelberg, Berlin, Heidelberg, 2014, pp. 67--122.

\bibitem{York:2010lsa}
R.~York, {FRIB: A New Accelerator Facility for the Production of and
  Experiments with Rare Isotope Beams}, in: {Particle Accelerator Conference
  (PAC 09)}, 2010, p. MO3GRI03.

\bibitem{Eschke_2005}
J.~Eschke, International facility for antiproton and ion research ({FAIR}) at
  {GSI}, darmstadt, Journal of Physics G: Nuclear and Particle Physics 31~(6)
  (2005) S967--S973.
\newblock \href {https://doi.org/10.1088/0954-3899/31/6/041}
  {\path{doi:10.1088/0954-3899/31/6/041}}.

\bibitem{BORGE2016408}
M.~Borge, Highlights of the isolde facility and the hie-isolde project, Nuclear
  Instruments and Methods in Physics Research Section B: Beam Interactions with
  Materials and Atoms 376 (2016) 408--412, proceedings of the XVIIth
  International Conference on Electromagnetic Isotope Separators and Related
  Topics (EMIS2015), Grand Rapids, MI, U.S.A., 11-15 May 2015.
\newblock \href {https://doi.org/https://doi.org/10.1016/j.nimb.2015.12.048}
  {\path{doi:https://doi.org/10.1016/j.nimb.2015.12.048}}.

\bibitem{Ball_2016}
G.~C. Ball, G.~Hackman, R.~Krücken, The {TRIUMF}-{ISAC} facility: two decades
  of discovery with rare isotope beams, Physica Scripta 91~(9) (2016) 093002.
\newblock \href {https://doi.org/10.1088/0031-8949/91/9/093002}
  {\path{doi:10.1088/0031-8949/91/9/093002}}.

\bibitem{KIM2020408}
Y.~J. Kim, Current status of experimental facilities at raon, Nuclear
  Instruments and Methods in Physics Research Section B: Beam Interactions with
  Materials and Atoms 463 (2020) 408--414.
\newblock \href {https://doi.org/https://doi.org/10.1016/j.nimb.2019.04.041}
  {\path{doi:https://doi.org/10.1016/j.nimb.2019.04.041}}.

\bibitem{WALLACE2007302}
M.~Wallace, M.~Famiano, M.-J. {van Goethem}, A.~Rogers, W.~Lynch, J.~Clifford,
  F.~Delaunay, J.~Lee, S.~Labostov, M.~Mocko, L.~Morris, A.~Moroni, B.~Nett,
  D.~Oostdyk, R.~Krishnasamy, M.~Tsang, R.~{de Souza}, S.~Hudan, L.~Sobotka,
  R.~Charity, J.~Elson, G.~Engel, The high resolution array (hira) for rare
  isotope beam experiments, Nuclear Instruments and Methods in Physics Research
  Section A: Accelerators, Spectrometers, Detectors and Associated Equipment
  583~(2) (2007) 302--312.
\newblock \href {https://doi.org/https://doi.org/10.1016/j.nima.2007.08.248}
  {\path{doi:https://doi.org/10.1016/j.nima.2007.08.248}}.

\bibitem{MUST2}
E.~Pollacco, D.~Beaumel, P.~Roussel-Chomaz, E.~Atkin, P.~Baron, J.~P. Baronick,
  E.~Becheva, Y.~Blumenfeld, A.~Boujrad, A.~Drouart, F.~Druillole,
  P.~Edelbruck, M.~Gelin, A.~Gillibert, C.~Houarner, V.~Lapoux, L.~Lavergne,
  G.~Leberthe, L.~Leterrier, V.~Le~Ven, F.~Lugiez, L.~Nalpas, L.~Olivier,
  B.~Paul, B.~Raine, A.~Richard, M.~Rouger, F.~Saillant, F.~Skaza, M.~Tripon,
  M.~Vilmay, E.~Wanlin, M.~Wittwer, Must2: A new generation array for direct
  reaction studies, in: C.~J. Gross, W.~Nazarewicz, K.~P. Rykaczewski (Eds.),
  The 4th International Conference on Exotic Nuclei and Atomic Masses, Springer
  Berlin Heidelberg, Berlin, Heidelberg, 2005, pp. 287--288.

\bibitem{WUOSMAA20071290}
A.~Wuosmaa, J.~Schiffer, B.~Back, C.~Lister, K.~Rehm, A solenoidal spectrometer
  for reactions in inverse kinematics, Nuclear Instruments and Methods in
  Physics Research Section A: Accelerators, Spectrometers, Detectors and
  Associated Equipment 580~(3) (2007) 1290--1300.
\newblock \href {https://doi.org/https://doi.org/10.1016/j.nima.2007.07.029}
  {\path{doi:https://doi.org/10.1016/j.nima.2007.07.029}}.

\bibitem{LIGHTHALL201097}
J.~Lighthall, B.~Back, S.~Baker, S.~Freeman, H.~Lee, B.~Kay, S.~Marley,
  K.~Rehm, J.~Rohrer, J.~Schiffer, D.~Shetty, A.~Vann, J.~Winkelbauer,
  A.~Wuosmaa, Commissioning of the helios spectrometer, Nuclear Instruments and
  Methods in Physics Research Section A: Accelerators, Spectrometers, Detectors
  and Associated Equipment 622~(1) (2010) 97--106.
\newblock \href {https://doi.org/https://doi.org/10.1016/j.nima.2010.06.220}
  {\path{doi:https://doi.org/10.1016/j.nima.2010.06.220}}.

\bibitem{Back2010}
B.~B. Back, S.~I. Baker, B.~A. Brown, C.~M. Deibel, S.~J. Freeman, B.~J.
  DiGiovine, C.~R. Hoffman, B.~P. Kay, H.~Y. Lee, J.~C. Lighthall, S.~T.
  Marley, R.~C. Pardo, K.~E. Rehm, J.~P. Schiffer, D.~V. Shetty, A.~W. Vann,
  J.~Winkelbauer, A.~H. Wuosmaa, First experiment with helios: The structure of
  $^{13}\mathbf{B}$, Phys. Rev. Lett. 104 (2010) 132501.
\newblock \href {https://doi.org/10.1103/PhysRevLett.104.132501}
  {\path{doi:10.1103/PhysRevLett.104.132501}}.

\bibitem{Santiago2018}
D.~Santiago-Gonzalez, K.~Auranen, M.~L. Avila, A.~D. Ayangeakaa, B.~B. Back,
  S.~Bottoni, M.~P. Carpenter, J.~Chen, C.~M. Deibel, A.~A. Hood, C.~R.
  Hoffman, R.~V.~F. Janssens, C.~L. Jiang, B.~P. Kay, S.~A. Kuvin, A.~Lauer,
  J.~P. Schiffer, J.~Sethi, R.~Talwar, I.~Wiedenh\"over, J.~Winkelbauer,
  S.~Zhu, Probing the single-particle character of rotational states in
  $^{19}\mathrm{F}$ using a short-lived isomeric beam, Phys. Rev. Lett. 120
  (2018) 122503.
\newblock \href {https://doi.org/10.1103/PhysRevLett.120.122503}
  {\path{doi:10.1103/PhysRevLett.120.122503}}.

\bibitem{Tang2020}
T.~L. Tang, B.~P. Kay, C.~R. Hoffman, J.~P. Schiffer, D.~K. Sharp, L.~P.
  Gaffney, S.~J. Freeman, M.~R. Mumpower, A.~Arokiaraj, E.~F. Baader, P.~A.
  Butler, W.~N. Catford, G.~de~Angelis, F.~Flavigny, M.~D. Gott, E.~T. Gregor,
  J.~Konki, M.~Labiche, I.~H. Lazarus, P.~T. MacGregor, I.~Martel, R.~D. Page,
  Z.~Podoly\'ak, O.~Poleshchuk, R.~Raabe, F.~Recchia, J.~F. Smith, S.~V. Szwec,
  J.~Yang, First exploration of neutron shell structure below lead and beyond
  $n=126$, Phys. Rev. Lett. 124 (2020) 062502.
\newblock \href {https://doi.org/10.1103/PhysRevLett.124.062502}
  {\path{doi:10.1103/PhysRevLett.124.062502}}.

\bibitem{SOLARISWP}
{SOLARIS, A Solenoidal Spectrometer Apparatus for Reaction Studies White
  Paper}, Tech. rep. (2018).

\bibitem{Ayyad2018}
{Ayyad, Y.}, {Bazin, D.}, {Beceiro-Novo, S.}, {Cortesi, M.}, {Mittig, W.},
  Physics and technology of time projection chambers as active targets, Eur.
  Phys. J. A 54~(10) (2018) 181.
\newblock \href {https://doi.org/10.1140/epja/i2018-12557-7}
  {\path{doi:10.1140/epja/i2018-12557-7}}.

\bibitem{BAZIN2020103790}
D.~Bazin, T.~Ahn, Y.~Ayyad, S.~Beceiro-Novo, A.~Macchiavelli, W.~Mittig,
  J.~Randhawa, Low energy nuclear physics with active targets and time
  projection chambers, Progress in Particle and Nuclear Physics 114 (2020)
  103790.
\newblock \href {https://doi.org/https://doi.org/10.1016/j.ppnp.2020.103790}
  {\path{doi:https://doi.org/10.1016/j.ppnp.2020.103790}}.

\bibitem{BECEIRONOVO2015124}
S.~Beceiro-Novo, T.~Ahn, D.~Bazin, W.~Mittig, Active targets for the study of
  nuclei far from stability, Progress in Particle and Nuclear Physics 84 (2015)
  124--165.
\newblock \href {https://doi.org/https://doi.org/10.1016/j.ppnp.2015.06.003}
  {\path{doi:https://doi.org/10.1016/j.ppnp.2015.06.003}}.

\bibitem{BRADT201765}
J.~Bradt, D.~Bazin, F.~Abu-Nimeh, T.~Ahn, Y.~Ayyad, S.~Beceiro Novo,
  L.~Carpenter, M.~Cortesi, M.~Kuchera, W.~Lynch, W.~Mittig, S.~Rost,
  N.~Watwood, J.~Yurkon, Commissioning of the active-target time projection
  chamber, Nuclear Instruments and Methods in Physics Research Section A:
  Accelerators, Spectrometers, Detectors and Associated Equipment 875 (2017)
  65--79.
\newblock \href {https://doi.org/https://doi.org/10.1016/j.nima.2017.09.013}
  {\path{doi:https://doi.org/10.1016/j.nima.2017.09.013}}.

\bibitem{POLESHCHUK2021165765}
O.~Poleshchuk, R.~Raabe, S.~Ceruti, A.~Ceulemans, H.~{De Witte}, T.~Marchi,
  A.~Mentana, J.~Refsgaard, J.~Yang, The specmat active target, Nuclear
  Instruments and Methods in Physics Research Section A: Accelerators,
  Spectrometers, Detectors and Associated Equipment 1015 (2021) 165765.
\newblock \href {https://doi.org/https://doi.org/10.1016/j.nima.2021.165765}
  {\path{doi:https://doi.org/10.1016/j.nima.2021.165765}}.

\bibitem{BRADT2018155}
J.~Bradt, Y.~Ayyad, D.~Bazin, W.~Mittig, T.~Ahn, S.~{Beceiro Novo}, B.~Brown,
  L.~Carpenter, M.~Cortesi, M.~Kuchera, W.~Lynch, S.~Rost, N.~Watwood,
  J.~Yurkon, J.~Barney, U.~Datta, J.~Estee, A.~Gillibert, J.~Manfredi,
  P.~Morfouace, D.~Pérez-Loureiro, E.~Pollacco, J.~Sammut, S.~Sweany, Study of
  spectroscopic factors at n=29 using isobaric analogue resonances in inverse
  kinematics, Physics Letters B 778 (2018) 155--160.
\newblock \href
  {https://doi.org/https://doi.org/10.1016/j.physletb.2018.01.015}
  {\path{doi:https://doi.org/10.1016/j.physletb.2018.01.015}}.

\bibitem{RANDHAWA2020}
J.~S. Randhawa, Y.~Ayyad, W.~Mittig, Z.~Meisel, T.~Ahn, S.~Aguilar,
  H.~Alvarez-Pol, D.~W. Bardayan, D.~Bazin, S.~Beceiro-Novo, D.~Blankstein,
  L.~Carpenter, M.~Cortesi, D.~Cortina-Gil, P.~Gastis, M.~Hall, S.~Henderson,
  J.~J. Kolata, T.~Mijatovic, F.~Ndayisabye, P.~O'Malley, J.~Pereira,
  A.~Pierre, H.~Robert, C.~Santamaria, H.~Schatz, J.~Smith, N.~Watwood, J.~C.
  Zamora, First direct measurement of
  $^{22}\mathrm{Mg}(\ensuremath{\alpha},p)^{25}\mathrm{Al}$ and implications
  for x-ray burst model-observation comparisons, Phys. Rev. Lett. 125 (2020)
  202701.
\newblock \href {https://doi.org/10.1103/PhysRevLett.125.202701}
  {\path{doi:10.1103/PhysRevLett.125.202701}}.

\bibitem{Fruhwirth429722}
R.~Frühwirth, M.~Regler, R.~K. Bock, H.~Grote, D.~Notz, M.~Regler,
  R.~Frühwirth, \href{https://cds.cern.ch/record/429722}{{Data analysis
  techniques for high-energy physics; 2nd ed.}}, Cambridge monographs on
  particle physics, nuclear physics, and cosmology, Cambridge Univ. Press,
  Cambridge, 2000.
\newline\urlprefix\url{https://cds.cern.ch/record/429722}

\bibitem{DALITZ2019159}
C.~Dalitz, Y.~Ayyad, J.~Wilberg, L.~Aymans, D.~Bazin, W.~Mittig, Automatic
  trajectory recognition in active target time projection chambers data by
  means of hierarchical clustering, Computer Physics Communications 235 (2019)
  159--168.
\newblock \href {https://doi.org/https://doi.org/10.1016/j.cpc.2018.09.010}
  {\path{doi:https://doi.org/10.1016/j.cpc.2018.09.010}}.

\bibitem{Kalman1960}
R.~E. Kalman, {A New Approach to Linear Filtering and Prediction Problems},
  Journal of Basic Engineering 82~(1) (1960) 35--45.
\newblock \href
  {http://arxiv.org/abs/https://asmedigitalcollection.asme.org/fluidsengineering/article-pdf/82/1/35/5518977/35\_1.pdf}
  {\path{arXiv:https://asmedigitalcollection.asme.org/fluidsengineering/article-pdf/82/1/35/5518977/35\_1.pdf}},
  \href {https://doi.org/10.1115/1.3662552} {\path{doi:10.1115/1.3662552}}.

\bibitem{FRUHWIRTH1987444}
R.~Frühwirth, Application of kalman filtering to track and vertex fitting,
  Nuclear Instruments and Methods in Physics Research Section A: Accelerators,
  Spectrometers, Detectors and Associated Equipment 262~(2) (1987) 444--450.
\newblock \href {https://doi.org/https://doi.org/10.1016/0168-9002(87)90887-4}
  {\path{doi:https://doi.org/10.1016/0168-9002(87)90887-4}}.

\bibitem{AYYAD2018166}
Y.~Ayyad, W.~Mittig, D.~Bazin, S.~Beceiro-Novo, M.~Cortesi, Novel particle
  tracking algorithm based on the random sample consensus model for the active
  target time projection chamber (at-tpc), Nuclear Instruments and Methods in
  Physics Research Section A: Accelerators, Spectrometers, Detectors and
  Associated Equipment 880 (2018) 166--173.
\newblock \href {https://doi.org/https://doi.org/10.1016/j.nima.2017.10.090}
  {\path{doi:https://doi.org/10.1016/j.nima.2017.10.090}}.

\bibitem{ZAMORA2021164899}
J.~Zamora, G.~Fortino, Tracking algorithms for tpcs using consensus-based
  robust estimators, Nuclear Instruments and Methods in Physics Research
  Section A: Accelerators, Spectrometers, Detectors and Associated Equipment
  988 (2021) 164899.
\newblock \href {https://doi.org/https://doi.org/10.1016/j.nima.2020.164899}
  {\path{doi:https://doi.org/10.1016/j.nima.2020.164899}}.

\bibitem{Cortesi2017}
M.~Cortesi, S.~Rost, W.~Mittig, Y.~Ayyad-Limonge, D.~Bazin, J.~Yurkon,
  A.~Stolz, Multi-layer thick gas electron multiplier (m-thgem): A new mpgd
  structure for high-gain operation at low-pressure, Review of Scientific
  Instruments 88~(1) (2017) 013303.
\newblock \href {https://doi.org/10.1063/1.4974333}
  {\path{doi:10.1063/1.4974333}}.

\bibitem{SUZUKI201239}
D.~Suzuki, M.~Ford, D.~Bazin, W.~Mittig, W.~Lynch, T.~Ahn, S.~Aune, E.~Galyaev,
  A.~Fritsch, J.~Gilbert, F.~Montes, A.~Shore, J.~Yurkon, J.~Kolata, J.~Browne,
  A.~Howard, A.~Roberts, X.~Tang, Prototype at-tpc: Toward a new generation
  active target time projection chamber for radioactive beam experiments,
  Nuclear Instruments and Methods in Physics Research Section A: Accelerators,
  Spectrometers, Detectors and Associated Equipment 691 (2012) 39--54.
\newblock \href {https://doi.org/https://doi.org/10.1016/j.nima.2012.06.050}
  {\path{doi:https://doi.org/10.1016/j.nima.2012.06.050}}.

\bibitem{Ayyad2019}
Y.~Ayyad, B.~Olaizola, W.~Mittig, G.~Potel, V.~Zelevinsky, M.~Horoi,
  S.~Beceiro-Novo, M.~Alcorta, C.~Andreoiu, T.~Ahn, M.~Anholm, L.~Atar,
  A.~Babu, D.~Bazin, N.~Bernier, S.~S. Bhattacharjee, M.~Bowry,
  R.~Caballero-Folch, M.~Cortesi, C.~Dalitz, E.~Dunling, A.~B. Garnsworthy,
  M.~Holl, B.~Kootte, K.~G. Leach, J.~S. Randhawa, Y.~Saito, C.~Santamaria,
  P.~\ifmmode \check{S}\else \v{S}\fi{}iuryt\ifmmode~\dot{e}\else \.{e}\fi{},
  C.~E. Svensson, R.~Umashankar, N.~Watwood, D.~Yates, Direct observation of
  proton emission in $^{11}\mathrm{Be}$, Phys. Rev. Lett. 123 (2019) 082501.
\newblock \href {https://doi.org/10.1103/PhysRevLett.123.082501}
  {\path{doi:10.1103/PhysRevLett.123.082501}}.

\bibitem{BILLOIR1985115}
P.~Billoir, R.~Frühwirth, M.~Regler, Track element merging strategy and vertex
  fitting in complex modular detectors, Nuclear Instruments and Methods in
  Physics Research Section A: Accelerators, Spectrometers, Detectors and
  Associated Equipment 241~(1) (1985) 115--131.
\newblock \href {https://doi.org/https://doi.org/10.1016/0168-9002(85)90523-6}
  {\path{doi:https://doi.org/10.1016/0168-9002(85)90523-6}}.

\bibitem{STAMPFER1994157}
D.~Stampfer, M.~Regler, R.~Frühwirth, Track fitting with energy loss, Computer
  Physics Communications 79~(2) (1994) 157--164.
\newblock \href {https://doi.org/https://doi.org/10.1016/0010-4655(94)90064-7}
  {\path{doi:https://doi.org/10.1016/0010-4655(94)90064-7}}.

\bibitem{FRU2000}
R.~Fruhwirth, M.~Regler, R.~Bock, H.~Grote, D.~Notz, Data Analysis Techniques
  for High-Energy Physics, Cambridge University Press, 2000.

\bibitem{HOPPNER2010518}
C.~Höppner, S.~Neubert, B.~Ketzer, S.~Paul, A novel generic framework for
  track fitting in complex detector systems, Nuclear Instruments and Methods in
  Physics Research Section A: Accelerators, Spectrometers, Detectors and
  Associated Equipment 620~(2) (2010) 518--525.
\newblock \href {https://doi.org/https://doi.org/10.1016/j.nima.2010.03.136}
  {\path{doi:https://doi.org/10.1016/j.nima.2010.03.136}}.

\bibitem{Daum2005}
F.~Daum, Nonlinear filters: beyond the kalman filter, IEEE Aerospace and
  Electronic Systems Magazine 20~(8) (2005) 57--69.
\newblock \href {https://doi.org/10.1109/MAES.2005.1499276}
  {\path{doi:10.1109/MAES.2005.1499276}}.

\bibitem{Rauch2021}
J.~Rauch, Dissertation, Technische Universität München, München (2021).

\bibitem{Shi2017}
{Noori Shirazi, Amir}, {Fleck, Ivor}, Bivariate normal distribution for finding
  inliers in hough space for a time projection chamber, EPJ Web Conf. 150
  (2017) 00010.
\newblock \href {https://doi.org/10.1051/epjconf/201715000010}
  {\path{doi:10.1051/epjconf/201715000010}}.

\bibitem{Ayyad_2017}
Y.~Ayyad, et~al., {Overview of the data analysis and new micro-pattern gas
  detector development for the Active Target Time Projection Chamber
  ({AT}-{TPC}) project.}, J. Phys. Conf. Ser. 876 (2017) 012003.
\newblock \href {https://doi.org/10.1088/1742-6596/876/1/012003}
  {\path{doi:10.1088/1742-6596/876/1/012003}}.

\bibitem{ATTPCROOT_git}
Y.~Ayyad, Attpcroot, \url{https://github.com/ATTPC/ATTPCROOTv2} (2023).

\bibitem{GIRAUD2023168213}
S.~Giraud, J.~Zamora, R.~Zegers, Y.~Ayyad, D.~Bazin, W.~Mittig, A.~Carls,
  M.~DeNudt, Z.~Rahman, Simulations and analysis tools for charge-exchange
  (d,2he) reactions in inverse kinematics with the at-tpc, Nuclear Instruments
  and Methods in Physics Research Section A: Accelerators, Spectrometers,
  Detectors and Associated Equipment (2023) 168213\href
  {https://doi.org/https://doi.org/10.1016/j.nima.2023.168213}
  {\path{doi:https://doi.org/10.1016/j.nima.2023.168213}}.

\bibitem{GIOMATARIS199629}
Y.~Giomataris, P.~Rebourgeard, J.~Robert, G.~Charpak, Micromegas: a
  high-granularity position-sensitive gaseous detector for high particle-flux
  environments, Nuclear Instruments and Methods in Physics Research Section A:
  Accelerators, Spectrometers, Detectors and Associated Equipment 376~(1)
  (1996) 29--35.
\newblock \href {https://doi.org/https://doi.org/10.1016/0168-9002(96)00175-1}
  {\path{doi:https://doi.org/10.1016/0168-9002(96)00175-1}}.

\bibitem{ZIEGLER20101818}
J.~F. Ziegler, et~al., {SRIM – The stopping and range of ions in matter
  (2010)}, Nucl. Instrum. Methods Phys. Res. B 268~(11) (2010) 1818--1823.
\newblock \href {https://doi.org/10.1016/j.nimb.2010.02.091}
  {\path{doi:10.1016/j.nimb.2010.02.091}}.

\bibitem{LOZOWSKI198954}
W.~Lozowski, Three diverse target preparations: 14c (12 mg/cm2), 71ga24mg (12
  mg/cm271ga, 3 mg/cm224mg), and 66,67zn (1.8–14.9 mg/cm2), Nuclear
  Instruments and Methods in Physics Research Section A: Accelerators,
  Spectrometers, Detectors and Associated Equipment 282~(1) (1989) 54--61.
\newblock \href {https://doi.org/https://doi.org/10.1016/0168-9002(89)90108-3}
  {\path{doi:https://doi.org/10.1016/0168-9002(89)90108-3}}.

\bibitem{POLLACCO201881}
E.~Pollacco, G.~Grinyer, F.~Abu-Nimeh, T.~Ahn, S.~Anvar, A.~Arokiaraj,
  Y.~Ayyad, H.~Baba, M.~Babo, P.~Baron, D.~Bazin, S.~Beceiro-Novo,
  C.~Belkhiria, M.~Blaizot, B.~Blank, J.~Bradt, G.~Cardella, L.~Carpenter,
  S.~Ceruti, E.~{De Filippo}, E.~Delagnes, S.~{De Luca}, H.~{De Witte},
  F.~Druillole, B.~Duclos, F.~Favela, A.~Fritsch, J.~Giovinazzo, C.~Gueye,
  T.~Isobe, P.~Hellmuth, C.~Huss, B.~Lachacinski, A.~Laffoley, G.~Lebertre,
  L.~Legeard, W.~Lynch, T.~Marchi, L.~Martina, C.~Maugeais, W.~Mittig,
  L.~Nalpas, E.~Pagano, J.~Pancin, O.~Poleshchuk, J.~Pedroza, J.~Pibernat,
  S.~Primault, R.~Raabe, B.~Raine, A.~Rebii, M.~Renaud, T.~Roger,
  P.~Roussel-Chomaz, P.~Russotto, G.~Saccà, F.~Saillant, P.~Sizun, D.~Suzuki,
  J.~Swartz, A.~Tizon, A.~Trifiró, N.~Usher, G.~Wittwer, J.~Yang, Get: A
  generic electronics system for tpcs and nuclear physics instrumentation,
  Nuclear Instruments and Methods in Physics Research Section A: Accelerators,
  Spectrometers, Detectors and Associated Equipment 887 (2018) 81--93.
\newblock \href {https://doi.org/https://doi.org/10.1016/j.nima.2018.01.020}
  {\path{doi:https://doi.org/10.1016/j.nima.2018.01.020}}.

\bibitem{Obertelli2014}
A.~Obertelli, A.~Delbart, S.~Anvar, L.~Audirac, G.~Authelet, H.~Baba,
  B.~Bruyneel, D.~Calvet, F.~Ch{\^a}teau, A.~Corsi, P.~Doornenbal, J.-M.
  Gheller, A.~Giganon, C.~Lahonde-Hamdoun, D.~Leboeuf, D.~Loiseau, A.~Mohamed,
  J.~P. Mols, H.~Otsu, C.~P{\'e}ron, A.~Peyaud, E.~C. Pollacco, G.~Prono, J.-Y.
  Rousse, C.~Santamaria, T.~Uesaka, Minos: A vertex tracker coupled to a thick
  liquid-hydrogen target for in-beam spectroscopy of exotic nuclei, The
  European Physical Journal A 50~(1) (2014) 8.
\newblock \href {https://doi.org/10.1140/epja/i2014-14008-y}
  {\path{doi:10.1140/epja/i2014-14008-y}}.

\bibitem{Macfarlane1978}
M.~H. Macfarlane, S.~C. Pieper, Ptolemy: a program for heavy-ion
  direct-reaction calculations, Tech. rep., United States, aNL--76-11(Rev1)
  (1978).

\bibitem{Chen2022}
J.~Chen, B.~P. Kay, T.~L. Tang, I.~A. Tolstukhin, C.~R. Hoffman, H.~Li, P.~Yin,
  X.~Zhao, P.~Maris, J.~P. Vary, G.~Li, J.~L. Lou, M.~L. Avila, Y.~Ayyad,
  S.~Bennett, D.~Bazin, J.~A. Clark, S.~J. Freeman, H.~Jayatissa,
  C.~M\"uller-Gatermann, A.~Munoz-Ramos, D.~Santiago-Gonzalez, D.~K. Sharp,
  A.~H. Wuosmaa, C.~X. Yuan, Probing the quadrupole transition strength of
  $^{15}\mathrm{C}$ via deuteron inelastic scattering, Phys. Rev. C 106 (2022)
  064312.
\newblock \href {https://doi.org/10.1103/PhysRevC.106.064312}
  {\path{doi:10.1103/PhysRevC.106.064312}}.

\end{thebibliography}





\end{document}